\documentclass[12pt,reqno]{amsart}

\usepackage{amsmath,amsfonts,amsthm,amssymb}

\usepackage[height=20cm,top=24mm,bottom=24mm,left=26mm,right=26mm]{geometry}

\usepackage[numbers,sort]{natbib}
\usepackage[]{hyperref}

\usepackage[]{graphicx}                                                    

\numberwithin{equation}{section}

\newcommand{\I}{\mathrm{i}}
\newcommand{\E}{\mathrm{e}}

\DeclareMathOperator{\sign}{sgn}
\DeclareMathOperator{\Tr}{Tr}

\DeclareMathOperator{\ch}{ch}
\DeclareMathOperator{\erfc}{erfc}

\DeclareMathDelimiter{\Norm}{\mathord}{largesymbols}{"3E}{largesymbols}{"3E}

\def\clap#1{\hbox to 0pt{\hss#1\hss}}

\allowdisplaybreaks[4]
\begin{document}
\baselineskip 16pt
\parskip 8pt
\sloppy

\begin{flushright}
YITP report 10-14
\end{flushright}

 \title[Superconformal Algebra and Entropy of Calabi--Yau
 Manifolds]{$\mathcal{N}=2$ Superconformal Algebra and the Entropy of
   Calabi--Yau Manifolds}


\author[T. Eguchi]{Tohru \textsc{Eguchi}}

\author[K. Hikami]{Kazuhiro \textsc{Hikami}}


\address{Yukawa Institute for Theoretical Physics, Kyoto University,
  Kyoto 606--8502, Japan}
\email{
  \texttt{eguchi@yukawa.kyoto-u.ac.jp}
}

\address{Department of Mathematics, 
  Naruto University of Education,
  Tokushima 772-8502, Japan.}

\email{
  \texttt{KHikami@gmail.com}
}

\date{\today}

\begin{abstract}
We use
the representation theory of
$\mathcal{N}=2$ superconformal algebra
to study the elliptic genera of Calabi--Yau (CY) $D$-folds.  
We compute the entropy of CY manifolds from the
growth rate of multiplicities
of the massive (non-BPS) representations in the 
decomposition of their elliptic genera.
We find that
the entropy of CY manifolds of complex dimension $D$ 
behaves differently depending on 
whether $D$ is even or odd.
When $D$ is odd, 
CY entropy  
coincides with the entropy of the corresponding 
hyperK{\"a}hler $(D-3)$-folds 
due to a structural theorem on Jacobi forms. 
In particular, we find that the Calabi--Yau $3$-fold 
has a vanishing entropy. At $D>3$, using our previous results on hyperK\"ahler 
manifolds, we find $S_{CY_D}\sim 2\pi \sqrt{{(D-3)^2\over 2(D-1)}n}$. When $D$ is even, we find the behavior of 
CY entropy behaving as
$S_{CY_D}\sim 2 \pi\sqrt{{D-1\over 2}n}$.
These agree with Cardy's formula at large $D$.
\end{abstract}





\maketitle
\section{Introduction}






The $\mathcal{N}=2$ superconformal algebra (SCA) 
is a basic tool in the world-sheet analysis of 
string compactifications on
Calabi--Yau (CY) manifold with complex dimension-$D$.
It is well-known that in $\mathcal{N}=2$ SCA there exist two types of
representations:
BPS (massless) and non-BPS (massive) representations.
BPS representations appear when the conformal weight $h$ of their
highest-weight state hits the unitarity 
bound $\frac{c}{24}=\frac{D}{8}$ (in the Ramond sector) where $c$
denotes the central charge of SCA. On the other hand non-BPS
representations appear at $h>\frac{D}{8}$.

We study the decomposition of the elliptic genus for 
Calabi--Yau manifold $CY_D$ in terms of characters of these representations;
\begin{multline}
  \text{elliptic genus of $CY_D$}
  =
  \sum_{Q}
  c_{D,Q} \,
  \left[
    \text{BPS representations with $U(1)$ charge-$Q$}
  \right]
  \\
  +
  \sum_{n=1}^\infty
  \sum_{Q}
  p_{D,Q}(n) \,
  \left[
    \text{non-BPS representations at $h=n +\frac{D}{8}$ with $U(1)$
      charge-$Q$}
  \right] .
\end{multline}
Since there exists only a finite number (of order $D$) of BPS
representations, the set of their multiplicities $c_{D,Q}$ is 
finite. On the other hand, we have an infinite series of
multiplicities $p_{D,Q}(n)$ for non-BPS representations and we define
the intrinsic CY entropy $S_{CY_D}$ by the rate of its exponential
growth 
\begin{equation}
  S_{CY_D} \sim \log  p_{D,Q}(n) .
\end{equation}

Such  an analysis has been done for the case of hyperK{\"a}hler
manifolds~\cite{EguchiHikami09a,EguchiHikami09b}
based on earlier works on
the representation theory of
the $\mathcal{N}=4$ superconformal
algebra~\cite{EgucTaor86a,EgucTaor88a,EgucTaor88b,EguOogTaoYan89a}.
It was pointed out 
that the predicted 
entropy of hyperK{\"a}hler manifolds coincides with that of the
standard $D1$-$D5$ black hole~\cite{StroVafa96a,CvetLars98a} 
when  we consider the symmetric product of $K3$ surfaces.
See  Ref.~\citenum{YokoNaka00a} for a similar idea. We also 
note that the expansion 
of elliptic genera in terms of theta functions has been discussed in   
Ref.~\citenum{DijMalMooVer00a}.

A key in our analysis is the fact that the characters of the BPS
representations  are the mock theta functions  
which were first introduced by Ramanujan 
(see, \emph{e.g.}, Refs.~\citenum{Dyson88walk,GEAndre89a,GordMcIn09a}).
We rely on recent developments~\cite{BrinKOno06a,Zweg02Thesis} on the
understanding of the mock theta function (see
Refs.~\citenum{Zagier08a,KOno08a} for review): namely, 
the  mock theta function
is a   holomorphic
part of the harmonic Maass form, and that it has a 
vector-valued modular form with weight-$3/2$ as its ``shadow''.

This paper is organized as follows.
In Section~\ref{sec:SCA} we  review
the characters of  the $\mathcal{N}=2$ superconformal algebra
and the elliptic genera of the CY manifolds.
We make an extensive use of Jacobi forms,
whose properties are collected in Appendix.
Decompositions of the elliptic genera for the CY 
manifold are studied in Sections~\ref{sec:CY_odd} and \ref{sec:CY_even}.
We treat the odd-dimensional CY manifolds in
Section~\ref{sec:CY_odd}.
We notice the close relationship between the characters of $\mathcal{N}=2$ and 
$\mathcal{N}=4$
SCA in this case and Gritsenko's result on the space of Jacobi forms.
It turns out that  
the entropy of CY $D$-folds coincides with the entropy of the corresponding 
hyperK{\"a}hler manifolds in $(D-3)$-dimensions. In particular the entropy of
CY 3-folds vanishes identically.
In   
Section~\ref{sec:CY_even} we discuss the even-dimensional CY manifolds.
Adopting the same strategy as  in our previous
paper~\cite{EguchiHikami09b}, we  compute the entropy by use of the
Poincar{\'e}--Maass series and obtain the result
\begin{equation*}
  S_{CY_D}\sim 2 \, \pi \, \sqrt{{D-1\over 2} \, n} .
\end{equation*}
The last section is devoted to concluding remarks.

\section{$\mathcal{N}=2$ Superconformal Algebras and
  Elliptic Genera}
\label{sec:SCA}

\subsection{$\mathcal{N}=2$ Superconformal Algebras}
The $\mathcal{N}=2$ superconformal algebra with the central
charge $c=3D$  
is a fundamental tool studying the compactification of string theory 
on Calabi-Yau manifold $CY_D$ with complex dimension $D$.
It is well-known that
there exists an isomorphism~\cite{SchwiSeibe87a} of the algebra,
\begin{equation}
  \label{spectral_flow}
  \begin{aligned}
    L_n
    & \to
    L_n + \alpha \, J_n + \frac{c}{6} \, \alpha^2 \, \delta_{n,0} ,
    \\[2mm]
    J_n
    & \to
    J_n + \frac{c}{3} \, \alpha \, \delta_{n,0} ,
    \\[2mm]
    G_r^{\pm}
    & \to
    G_{r\pm \alpha}^\pm ,
  \end{aligned}
\end{equation}
where $\alpha \in \mathbb{Z}/2$. 
Ramond and NS sectors are exchanged when 
$\alpha\in\mathbb{Z}+\frac{1}{2}$.
For our purpose of studying string compactification, we
need  the extended 
$\mathcal{N}=2$ SCA which is invariant under the integral spectral
flow
$\alpha\in\mathbb{Z}$.
Odake discussed
such series of  extended SCA's 
by adding spectral flow generators to the standard $\mathcal{N}=2$
SCA~\cite{Odake89a,Odake90a,Odake90b}.

The highest weight state in the extended $\mathcal{N}=2$ SCA are
labeled by conformal weight~$h$ and $U(1)$ charge~$Q$;
\begin{equation}
  \begin{gathered}
    L_0  \left| \Omega \right\rangle
    =
    h  \left| \Omega \right\rangle ,
    \\[2mm]
    J_0   \left| \Omega \right\rangle
    =
    Q  \left| \Omega \right\rangle .
  \end{gathered}
\end{equation}
In the Ramond sector, due to a relation between zero-modes of the
supercurrents,
$\{ G_0^+, G_0^-\} = 2 \,\left( L_0 - \frac{c}{24}\right)$, 
we have the unitarity condition 
\begin{equation}
  h  \geq
  \frac{D}{8}
  .
\end{equation}
Characters are defined by
\begin{equation}
  \label{define_character}
  \ch_{D,h,Q}^* (z; \tau)
  =
  \Tr_{\mathcal{H}^*}
  \left(
    q^{L_0 - \frac{c}{24}} \,
    \E^{2 \pi \I z J_0}
  \right) ,
\end{equation}
where $*$ denotes the spin structure, and $\mathcal{H}^*$ is the
Hilbert space of the representation.
Under the spectral flow~\eqref{spectral_flow},
the characters in  the Ramond/NS sectors
are transformed to each other as;
\begin{equation}
  \begin{gathered}
    \ch_\bullet^{\widetilde{NS}}(z;\tau)
    =
    \ch_\bullet^{NS} \left( z  + \frac{1}{2} ; \tau \right) ,
    \\[2mm]
    \ch_\bullet^{R}(z;\tau)
    =
    q^{\frac{D}{8}} \, \E^{D \pi \I z} \,
    \ch_\bullet^{NS} \left( z + \frac{\tau}{2}; \tau \right) ,
    \\[2mm]
    \ch_\bullet^{\widetilde{R}}(z;\tau)
    =
    \E^{  \frac{D}{2} \pi \I} \,
    \ch_\bullet^{R} \left( z + \frac{1}{2}; \tau \right) .
  \end{gathered}
\end{equation}
Here the phase factor in the
$\widetilde{R}$-sector is our convention.

By construction,
the superconformal character~\eqref{define_character}
is given by the irreducible characters of $\mathcal{N}=2$
SCA~\cite{Kirit88b,Dobre87a} summed over spectral flow.
This fact can be directly checked
in the work of Refs.~\citenum{Odake90a,Odake90b}.

In the Ramond sector, characters are given explicitly as follows.
Here the $U(1)$ charge takes values
\begin{equation}
  Q \equiv \frac{D}{2} \mod \mathbb{Z}.
\end{equation}
\begin{itemize}
\item massive (non-BPS) representations:\\
 $h> \frac{D}{8}$;
  $Q=\frac{D}{2}, \frac{D}{2} -1, \dots,-(\frac{D}{2}-1),
  -\frac{D}{2}$ and $Q\neq 0 \,(D=\mbox{even}) $,
  \begin{multline}
    \label{massive_character}
    \ch_{D,h,Q > 0}^{\widetilde{R}}(z;\tau)
    =
    (-1)^{Q+ \frac{D}{2} - 1} \, q^{h - \frac{D}{8}} \,
    \frac{
      \I \, \theta_{11}(z;\tau)}{
      \left[ \eta(\tau) \right]^3
    } \,
    \E^{2 \pi \I \left( Q - \frac{1}{2} \right) z}
    \\
    \times
    \sum_{n \in \mathbb{Z}}
    q^{\frac{D-1}{2} n^2 + \left( Q - \frac{1}{2} \right) n} \,
    \left(
      - \E^{2 \pi \I z}
    \right)^{(D-1) n} ,
  \end{multline}    

\item massless (BPS) representations:\\
 $h=\frac{D}{8}$;
  $Q=\frac{D}{2}-1, \frac{D}{2}-2,
  \dots,
  - \left( \frac{D}{2} -1 \right)$,
  \begin{multline}
    \label{massless_character}
    \ch_{D,h=\frac{D}{8},Q\geq 0}^{\widetilde{R}}(z;\tau)
    =
    (-1)^{Q+ \frac{D}{2} } \,
    \frac{\I \, \theta_{11}(z;\tau)}{
      \left[ \eta(\tau) \right]^3} \,
    \E^{2 \pi \I \left( Q  + \frac{1}{2} \right) z }
    \\
    \times
    \sum_{n \in \mathbb{Z}}
    q^{\frac{D-1}{2} n^2 + \left( Q+\frac{1}{2} \right) n} \,
    \frac{
      \left( - \E^{2 \pi \I z} \right)^{(D-1) n}
    }{
      1 - \E^{2 \pi \I z} \, q^n
    } ,
  \end{multline}
  and for $h=\frac{D}{8}$; $Q=\frac{D}{2}$
  \begin{multline}
    \ch^{\widetilde{R}}_{D,h=\frac{D}{8},Q=\frac{D}{2}}(z;\tau)
    =
    (-1)^D \,
    \frac{
      \I \, \theta_{11}(z;\tau)}{
      \left[ \eta(\tau) \right]^3} \,
    \E^{2 \pi \I  \frac{D+1}{2} z}
    \\
    \times
    \sum_{n \in \mathbb{Z}}
    q^{\frac{D-1}{2} n^2 + \frac{D+1}{2} n}
    \,
    \frac{
      \left( 1-q \right) \,
      \left( 
        - \E^{2 \pi \I z}
      \right)^{(D-1) n} 
    }{
      \left( 1 - \E^{2 \pi \I z} \, q^n
      \right) \,
      \left( 1 - \E^{2 \pi \I z} \, q^{n+1}
      \right)
    } .
  \end{multline}
\end{itemize}
See Appendix~\ref{sec:Theta} for the notation of theta functions.
The characters for $Q<0$ are given by
\begin{equation}
  \ch_{D,h,- Q < 0}^{\widetilde{R}}(z;\tau)
  =
  \ch_{D,h,Q}^{\widetilde{R}}(-z ; \tau) .
\end{equation}

The Witten index of massless representations are given by
\begin{equation}
  \ch^{\widetilde{R}}_{D,h = \frac{D}{8}, Q \geq 0 }(z=0; \tau)
  =
  \begin{cases}
    (-1)^{Q+\frac{D}{2}} ,
    & \text{for $0 \leq Q < \frac{D}{2}$,}
    \\[2mm]
    1+ (-1)^D ,
    & \text{for $Q=\frac{D}{2}$,}
  \end{cases}
\end{equation}
while all massive representations have a vanishing index.

At the unitarity bound $h=\frac{D}{8}$, a massive character
decomposes into a sum of massless characters as
\begin{equation}
  \label{unitary_bound_Q}
  \lim_{h\searrow \frac{D}{8}} \ch_{D,h,Q+1}^{\widetilde{R}}(z;\tau)
  =
  \ch_{D,h=\frac{D}{8},Q+1}^{\widetilde{R}}(z;\tau)
  +
  \ch_{D,h=\frac{D}{8},Q}^{\widetilde{R}}(z;\tau)
  ,
\end{equation}
where $Q \geq 0$,
and
\begin{multline}
  \label{unitary_bound_zero}
  \lim_{h \searrow \frac{D}{8}}
    \ch_{D,h,Q=\frac{D}{2}}^{\widetilde{R}}(z;\tau)
  \\
  =
  \ch_{D, h=\frac{D}{8}, Q=\frac{D}{2}}^{\widetilde{R}}(z;\tau)
  +
  \ch_{D,h=\frac{D}{8},Q=\frac{D}{2}-1}^{\widetilde{R}}(z;\tau)
  +
  \ch_{D,h=\frac{D}{8},Q=- \left( \frac{D}{2}-1 \right)}^{\widetilde{R}}(z;\tau)
  .
\end{multline}

\subsection{Elliptic Genus}
The elliptic genus of the CY manifold with complex
dimension $D$ is identified with~\cite{EWitt87a}
\begin{equation}
  Z_{CY_D}(z;\tau)
  =
  \Tr_{\mathcal{H}^R \otimes \mathcal{H}^R}
  \left[
    (-1)^F \,
    \E^{2 \pi \I z J_0} \,
    q^{L_0 - \frac{D}{8}} \,
    \overline{q}^{\overline{L}_0 - \frac{D}{8}}
  \right] ,
\end{equation}
where
$(-1)^F=\E^{\pi \I \left( J_0 - \overline{J}_0 \right)}$.
Due to the supersymmetry, 
only the ground state contributes in the right-moving sector, and
the elliptic genus is independent of
$\overline{q}$.
It is known~\cite{KawaYamaYang94a} that the elliptic genus is the
Jacobi form with weight-$0$ and index-$\frac{D}{2}$.
See Appendix~\ref{sec:Jacobi} for the definition of the Jacobi form.

The elliptic genus is related to the topological invariants of
manifolds.
We have
\begin{equation}
  \begin{gathered}
    Z_{CY_D}(z=0;\tau) =
    \chi_{CY_D},
    \\[2mm]
%
    q^{\frac{D}{4}} \,
    Z_{CY_D} \left(
      z=\frac{1+\tau}{2}; \tau
      \right)
      =
      \widehat{A}_{CY_D}+\cdots ,
  \end{gathered}
\end{equation}
where $\chi_{CY_D}$ and $\widehat{A}_{CY_D}$ are
respectively the Euler characteristic and
the $\widehat{A}$-genus.
\section{Calabi--Yau Manifolds:  Odd-Dimension}
\label{sec:CY_odd}
We study odd-dimensional Calabi-Yau manifolds $CY_D$ 
($D=$ odd) 
throughout this section.
The $U(1)$ charge is
$Q\in \mathbb{Z}+\frac{1}{2}$.

\subsection{Character Decomposition}

We see from~\eqref{massless_character} that the combination of 
$\mathcal{N}=2$ massless characters which is even in $z$ can
be written as
\begin{equation}
  \label{massless_and_C_k}
  \begin{aligned}[b]
    & \ch_{D,h=\frac{D}{8},Q=\frac{1}{2}}^{\widetilde{R}}(z;\tau)
    +
    \ch_{D,h=\frac{D}{8},Q=-\frac{1}{2}}^{\widetilde{R}}(z;\tau)
    \\
    & = 
    (-1)^{\frac{D+1}{2}} \,
    \frac{\I \, \theta_{11}(z;\tau)}{
      \left[
        \eta(\tau)
      \right]^3
    } \,
    \sum_{n \in \mathbb{Z}}
    q^{\frac{D-1}{2} n^2} \,
    \E^{2 \pi \I (D-1) n z} \,
    \frac{
      1+ \E^{2 \pi \I z} \, q^n
    }{
      1 - \E^{2 \pi \I z} \, q^n
    } 
    \\
    &= 
    (-1)^{\frac{D+1}{2}} \,
    \phi_{0,\frac{3}{2}}(z;\tau) \,
    C^{\mathcal{N}=4}_{\frac{D-3}{2}}(z;\tau) ,
  \end{aligned}
\end{equation}
where $C^{\mathcal{N}=4}_k(z;\tau)$, 
defined in~\eqref{define_C_k}~\cite{EguchiHikami09b}, is
the
isospin-$0$ massless character in
$c=6 \, k$ $\mathcal{N}=4$
SCA~\cite{EgucTaor86a,EgucTaor88b,EgucTaor88a}.
$\phi_{0,\frac{3}{2}}(z,\tau)$
is a  weight-$0$ index-${3/2}$ Jacobi form listed in
Appendix~\ref{sec:Jacobi}. 
Similarly the even-$z$ combination of  massive
characters~\eqref{massive_character} can be written as
\begin{equation}
  \label{massive_and_B_k_a}
  \ch_{D,h,Q}^{\widetilde{R}}(z;\tau)
  +
  \ch_{D,h,-Q}^{\widetilde{R}}(z;\tau)
  =
  (-1)^{Q+\frac{D}{2}-1} \,
  \phi_{0,\frac{3}{2}}(z;\tau) \,
  q^{h - \frac{D}{8} - \frac{\left(Q-\frac{1}{2}\right)^2}{2 (D-1)}}
  \,
  B^{\mathcal{N}=4}_{\frac{D-3}{2}, Q - \frac{1}{2}}(z;\tau) ,
\end{equation}
where $Q>0$, and
$B^{\mathcal{N}=4}_{k,a}(z;\tau)$, defined
in~\eqref{define_B_k_a}, is the basis of massive characters in
$\mathcal{N}=4$ SCA~\cite{EguchiHikami09b}.
So both the (even-$z$ part of) $\mathcal{N}=2$ massless and massive
characters  coincide with those of
$\mathcal{N}=4$ SCA up to the Jacobi form
$\phi_{0,\frac{3}{2}}(z;\tau)$.

This also happens to the elliptic genus $Z_{CY_D}(z;\tau)$.  
The elliptic genus for $CY_D$ is a Jacobi form with weight-$0$ and
index-$\frac{D}{2}$ and is an even function of $z$ because
of~\eqref{Jacobi-def}.
Now 
there exists a structural theorem on the space of Jacobi forms with
half-integral index
$\mathbb{J}_{0,\frac{D}{2} \in  \mathbb{Z}+\frac{1}{2}}$~\cite{Gritse99a}. 
This space is isomorphic to the space  
$\mathbb{J}_{0,\frac{D-3}{2} \in \mathbb{Z}}$
\begin{equation}
  \mathbb{J}_{0,\frac{D}{2}} 
  =
  \phi_{0,\frac{3}{2}}(z;\tau) \cdot
  \mathbb{J}_{0,\frac{D-3}{2}} .
  \label{Gritsenko}
\end{equation}
In view of the relationship~\eqref{massless_and_C_k},~\eqref{massive_and_B_k_a} and the above theorem~\eqref{Gritsenko},
we can conclude that the character decomposition of the elliptic genus
of the Calabi--Yau $D$-fold
is essentially the same as  
that of the hyperK\"{a}hler manifolds with
complex dimension-$(D-3)$, which was 
studied in
our previous paper~\cite{EguchiHikami09b}.

Let us consider the case when 
the elliptic genus $Z_{CY_D}(z;\tau)$ includes a piece
\begin{equation}
  \phi_{0,\frac{3}{2}}(z;\tau)
  \,
  \left[
    \left(
      \frac{
        \theta_{10}(z;\tau)}{
        \theta_{10}(0;\tau)}
    \right)^{D-3}
    +
    \left(
      \frac{
        \theta_{00}(z;\tau)}{
        \theta_{00}(0;\tau)}
    \right)^{D-3}
    +
    \left(
      \frac{
        \theta_{01}(z;\tau)}{
        \theta_{01}(0;\tau)}
    \right)^{D-3}
  \right] .
  \label{dominant}
\end{equation}
The above combination 
contains the identity representation in the NS sector 
and gives the dominant contribution to the entropy. 
Using the results of Ref.~\citenum{EguchiHikami09b} we can derive  
the entropy of CY manifolds $S_{CY_D}$ from
the increase in the multiplicity of massive representations
with $U(1)$ charge $Q>0$
\begin{equation}
  S_{CY_D}
  \sim
  2 \, \pi \,
  \sqrt{
    \frac{(D-3)^2}{2 \, (D-1)} \,
    n
    -
    \left(
      \frac{D-3}{D-1} \, 
      \frac{Q-\frac{1}{2}}{2}
    \right)^2
  } .
\end{equation}
\subsection{Examples}
\subsubsection{Calabi--Yau $3$-fold $CY_3$}
The elliptic genus of the Calabi--Yau $3$-fold $CY_3$ is given
by~\cite{KawaYamaYang94a,CNeum99a,Gritse99a}
\begin{equation}
  Z_{CY_3}(z;\tau)
  = \frac{\chi_{CY_3}}{2} \, \phi_{0,\frac{3}{2}}(z;\tau) ,
\end{equation}
where
$\chi_{CY_3}$ is the Euler number of $CY_3$.

With a help of~\eqref{massless_and_C_k} we have
\begin{equation*}
  \ch_{D=3,h=\frac{3}{8}, Q=\frac{1}{2}}^{\widetilde{R}}(z;\tau)
  +
  \ch_{D=3,h=\frac{3}{8}, Q=-\frac{1}{2}}^{\widetilde{R}}(z;\tau)
  =
  \phi_{0,\frac{3}{2}}(z;\tau) ,
\end{equation*}
which proves a simple decomposition formula for the elliptic genus,
\begin{equation}
  Z_{CY_3}(z;\tau)
  =
  \frac{\chi_{CY_3}}{2} \,
  \left[
    \ch_{D=3,h=\frac{3}{8}, Q=\frac{1}{2}}^{\widetilde{R}}(z;\tau)
    +
    \ch_{D=3,h=\frac{3}{8}, Q=-\frac{1}{2}}^{\widetilde{R}}(z;\tau)
  \right] .
\end{equation}
There is no contribution from the massive representations in this
decomposition,
and we 
conclude that the entropy of the CY 3-fold vanishes identically,
\begin{equation}
  S_{CY_3} = 0.
\end{equation}

\subsubsection{Calabi--Yau $5$-fold $CY_5$}
The elliptic genus of the Calabi--Yau $5$-fold $CY_5$ is
\begin{equation}
  \label{Z_CY5}
  Z_{CY_5}(z;\tau)
  = \frac{\chi_{CY_5}}{24} \,
  \phi_{0,\frac{3}{2}}(z;\tau) \,
  \phi_{0,1}(z;\tau) ,
\end{equation}
where $\chi_{CY_5}$ is the Euler number of $CY_5$ and $\phi_{0,1}$ is
one half of the elliptic genus of $K3$ surface~\eqref{phi01}.
As was shown in Ref.~\citenum{EguchiHikami09a} (see also
Ref.~\citenum{EguOogTaoYan89a}),
character decomposition of the $K3$ elliptic genus is  given by 
\begin{equation}
  \label{decomposition_K3}
  2 \, \phi_{0,1}(z;\tau)
  =
  24 \, C^{\mathcal{N}=4}_1(z;\tau)
  - 
  q^{-\frac{1}{8}} \,
  \left[
    2 - \sum_{n=1}^\infty A_n \, q^n
  \right] \,
  B^{\mathcal{N}=4}_{1,1}(z;\tau) .
\end{equation}  
Here the
 positive integers $A_n$ are given by the Rademacher expansion as
\begin{equation}
  \label{K3_An}
  A_n =
  \frac{- 2 \, \pi \, \I}{
    \left( 8 \, n - 1\right)^{\frac{1}{4}}
  }
  \sum_{c=1}^\infty
  \frac{1}{\sqrt{c}} \,
  I_{\frac{1}{2}}
  \left(
    \frac{\pi \, \sqrt{ 8 \, n -1}}{2 \, c}
  \right) \,
  \sum_{\substack{
      k \mod 4 c
      \\
      k^2= - 8 n +1 \mod 8c
    }}
  \left(
    \frac{-4}{k}
  \right) \,
  \E^{\frac{k}{2 c} \pi \I} ,
\end{equation}
where
$\left(\frac{-4}{\bullet}\right)$ is the Legendre symbol,
and $I_k(x)$ denotes the modified Bessel function,
\begin{equation}
  I_{\frac{1}{2}}(x)
  =
  \sqrt{\frac{2}{\pi \, x}} \,
  \sinh(x) .
\end{equation}
Since $\mathcal{N}=2$ characters coincide with those of
$\mathcal{N}=4$ up to a factor $\phi_{0,{3\over 2}}$,
the character decomposition of~\eqref{Z_CY5} becomes exactly the same
as in~\eqref{decomposition_K3}.
We conclude that
\begin{equation}
  \begin{aligned}[b]
    S_{CY_5}
    & = S_{K3}=
    \log A_n \\
    &   \sim
    2 \, \pi
    \sqrt{
      \frac{1}{2} \,
      \left(
        n - \frac{1}{8}
      \right)
    } .
  \end{aligned}
\end{equation}

\section{Calabi--Yau Manifolds:  Even-Dimension}
\label{sec:CY_even}
Let us next study the even-dimensional Calabi--Yau manifolds $CY_D$.
$D$ is 
even 
throughout this section.
The $U(1)$ charge is integral 
$Q\in \mathbb{Z}$.
We follow the method of Ref.~\citenum{EguchiHikami09b} to 
construct the character decomposition of the elliptic genus.

\subsection{Massless and Massive Characters}
For notational simplicity we set ${C}_D(z;\tau)$ as the massless character
with $U(1)$ charge-$0$,
\begin{equation}
  \begin{aligned}[b]
    {C}_D(z;\tau)
    & =
    (-1)^{\frac{D}{2}} \,
    \ch_{D, h=\frac{D}{8}, Q=0}^{\widetilde{R}}(z;\tau)
    \\
    & =
    \frac{\I \, \theta_{11}(z;\tau)}{
      \left[
        \eta(\tau)
      \right]^3
    } \,
    \E^{\pi \I z}
    \sum_{n \in \mathbb{Z}}
    (-1)^n \,
    q^{\frac{D-1}{2} n^2 + \frac{1}{2} n} \,
    \frac{
      \E^{2 \pi \I (D-1)n z}
    }{
      1 - \E^{2 \pi \I z} \, q^n
    } .
  \end{aligned}
\end{equation}
It is identified as 
\begin{equation}
  {C}_D(z;\tau)
  =
  \frac{
    \I \, \theta_{11}(z;\tau)}{
    \left[
      \eta(\tau) \right]^3} \,
  \E^{\pi \I z} \,
  f_{D-1}
  \left(
    \frac{-1+\tau}{2 \, (D-1)} ,
    z+ \frac{-1+\tau}{2 \, (D-1)}
    ;
    \tau
  \right) ,
\end{equation}
where $f_D(u,z;\tau)$ is the Appell function
defined in~\eqref{define_f}. Appell function undergoes a modular transformation with a  Mordell's integral (\ref{Mordell-int}) and is thus a typical mock theta function. One can cure its modular property by the process of completion~\cite{Zweg02Thesis}. 
By using~\eqref{completion_f},
we define the
completion
\begin{multline}
  \widehat{{C}}_D(z;\tau)
  =
  {C}_D(z;\tau)
  \\
  -
  \frac{1}{2} \,
  \sum_{a \mod (D-1)}
  \left[
    \E^{- \frac{a-1}{D-1}\pi \I} \,
    q^{-\frac{1}{8 (D-1)}} \,
    R_{\frac{D-1}{2},a-1} \left(
      \frac{-1+\tau}{2 \, (D-1)}; \tau
    \right)
  \right] \,
  \phi_{-1,\frac{1}{2}}(z;\tau) \,
  \widetilde{\vartheta}_{\frac{D-1}{2},a}(z;\tau) ,
\end{multline}
where $\widetilde{\vartheta}_{\frac{D-1}{2},a}(z;\tau)$ is a (modified) theta
series defined as~\eqref{define_Big_theta},
and
$R_{{D\over 2},a}(z;\tau)$ denotes
the non-holomorphic function~\eqref{R-func}.
(Note that the combination
$\E^{-{\frac{a-1}{D-1}\pi  \I}} \,
R_{{D-1\over 2},a-1}\left(
  \frac{-1+\tau}{2(D-1)}; \tau
\right)$ is invariant
under $a\rightarrow D-a$).

One finds that $\widehat{{C}}_D(z;\tau)$
is
a real analytic Jacobi form satisfying
\begin{equation}
  \begin{gathered}
    \widehat{{C}}_D \left(
      \frac{z}{\tau} ; - \frac{1}{\tau}
    \right)
    =
    \E^{\pi \I D \frac{z^2}{\tau}} \,
    \widehat{{C}}_D(z;\tau),
    \\[2mm]
    \widehat{{C}}_D(z+\tau; \tau)
    =
    q^{-\frac{D}{2}} \, \E^{- 2 \pi \I D z} \,
    \widehat{{C}}_D(z;\tau) ,
    \\[2mm]
    \widehat{{C}}_D(z;\tau+1)
    =
    \widehat{{C}}_D(z+1;\tau)
    =
    \widehat{{C}}_D(z;\tau) .
  \end{gathered}
\end{equation}
We note that the shadow~\cite{Zagier08a},
a weight-$3/2$ vector modular form,
of our system of mock theta functions  is given by
\begin{equation}
  \label{differential_R}
  \begin{aligned}[b]
    & \I \, \sqrt{2 \, (D-1)} \, 
    \overline{
      \sqrt{\Im \tau} \,
      \frac{\partial}{\partial \overline{\tau}}
      \left[
        \E^{- \frac{a-1}{D-1} \pi \I} \, q^{-\frac{1}{8(D-1)}} \,
        R_{\frac{D-1}{2},a-1}
        \left(
          \frac{-1+\tau}{2(D-1)} ; \tau
        \right)
      \right]
    }
    \\
    & =
    \sum_{n \in \mathbb{Z}}
    \left(
      (D-1) \, n + a-\frac{1}{2}
    \right) \,(-1)^n \,
    q^{\frac{1}{2 (D-1)}
      \left(
        (D-1) n + a - \frac{1}{2}
      \right)^2
    }
    \\
    & =
    \frac{1}{2}
    \left[
      \eta(\tau)
    \right]^3 \,
    \frac{
      \widetilde{\vartheta}_{\frac{D-1}{2},a}
      + \widetilde{\vartheta}_{\frac{D-1}{2},D-a}
    }{
      \widetilde{\vartheta}_{\frac{1}{2},1}
    } ( 0 ; \tau) ,
  \end{aligned}
\end{equation}
where
$1 \leq a \leq \frac{D}{2}  $.

The theta series
$\widetilde{\vartheta}_{\frac{D-1}{2},a}(\tau)$~\eqref{define_Big_theta},
which is a basis for
Jacobi
form with index $\frac{D-1}{2}\in
\mathbb{Z}+\frac{1}{2}$,
may be regarded as
the massive characters~\eqref{massive_character} up to a prefactor;
\begin{equation}
  \label{massive_and_Theta}
  \ch_{D,h> \frac{D}{8},Q}^{\widetilde{R}}(z;\tau)
  =
  (-1)^{Q+\frac{D}{2}-1} \,
  q^{h - \frac{D}{8}- \frac{\left( Q - \frac{1}{2}\right)^2}{2(D-1)}} \,
  \phi_{-1,\frac{1}{2}}(z;\tau) \,
  \widetilde{\vartheta}_{\frac{D-1}{2},Q}(z;\tau) .
\end{equation}
Combining the theta series,
we define the basis functions as
\begin{equation}
  {B}_{D,a}(z;\tau)
  =
  \begin{cases}
    \displaystyle
    \phi_{-1,\frac{1}{2}}(z;\tau) \,
    \left[
      \widetilde{\vartheta}_{\frac{D-1}{2},a}(z;\tau)
      +
      \widetilde{\vartheta}_{\frac{D-1}{2},D-a}(z;\tau)
    \right] ,
    &\text{for $1 \leq a < \frac{D}{2}$},
    \\[6mm]
    \displaystyle
    \phi_{-1,\frac{1}{2}}(z;\tau) \,
    \widetilde{\vartheta}_{\frac{D-1}{2},\frac{D}{2}}(z;\tau)
    ,
    &\text{for $a = \frac{D}{2} $.}
  \end{cases}
\end{equation}
Note that
$
\widetilde{\vartheta}_{\frac{D-1}{2},D-a}(z;\tau)
=-
\widetilde{\vartheta}_{\frac{D-1}{2},a}(-z;\tau)$
and $B_{D,a}(z;\tau)$ is an even-function
of $z$ with a 2nd order zero at $z=0$.
One has 
\begin{equation}
  \ch_{D,h> \frac{D}{8},Q}^{\widetilde{R}}(z;\tau)+
   \ch_{D,h> \frac{D}{8},-Q}^{\widetilde{R}}(z;\tau)  =
  (-1)^{Q+\frac{D}{2}-1} \,
  q^{h - \frac{D}{8}- \frac{\left( Q - \frac{1}{2}\right)^2}{2(D-1)}}
  \,
  B_{D,Q}(z;\tau) .  
\end{equation}
Thus they describe the even-$z$ part of the massive characters.
   
They form a set of vector-valued Jacobi forms satisfying
\begin{equation}
  \begin{gathered}[b]
    {B}_{D,a}(z+1;\tau)
    = {B}_{D,a}(z;\tau) ,
    \\[2mm]
    {B}_{D,a}(z+\tau; \tau)
    =
    q^{- \frac{D}{2}} \,
    \E^{- 2 \pi \I D z} \,
    {B}_{D,a}(z;\tau) ,
    \\[2mm]
    {B}_{D,a}(z;\tau)
    =
    \sqrt{\frac{\tau}{\I}} \,
    \E^{- \pi \I D \frac{z^2}{\tau}} \,
    \sum_{b=1}^{\frac{D}{2}} 
    \frac{ \delta_{a, \frac{D}{2} } -2}{\sqrt{D-1}} \,
    \sin \left(
      \frac{(2 \, a - 1) \, (2 \, b - 1)}{2 \, (D-1)} \,
      \pi \right)
    \,
    {B}_{D,b} \left( \frac{z}{\tau} ; - \frac{1}{\tau}
    \right) ,
    \\[2mm]
    {B}_{D,a}(z;\tau+1)
    =
    \E^{
      \frac{\left( a - \frac{1}{2} \right)^2}{D-1} \pi \I} \,
    {B}_{D,a}(z;\tau) .
  \end{gathered}
\end{equation}
By using ~\eqref{modular_Theta},
the latter two identities
are summarized  as
\begin{multline}
  \label{general_modular_B}
  {B}_{D,a_1}
  \left(
    \frac{z}{c \, \tau+d} ;
    \frac{a \, \tau +b}{c \, \tau + d}
  \right)
  \\
  =
  \begin{cases}
    \displaystyle
    \frac{1}{\sqrt{c \, \tau+d}} \,
    \E^{D  \frac{c z^2}{c \tau+d}  \pi \I}
    \sum_{a_2=1}^{\frac{D}{2}}
    \left(
      \left[
        \rho(\gamma)
      \right]_{a_1,a_2}
      +
      \left[
        \rho(\gamma)
      \right]_{D-a_1,a_2} 
    \right) \,
    {B}_{D,a_2}(z;\tau) ,
    & 
    \text{for $a_1<\frac{D}{2} $,}
    \\[6mm]
    \displaystyle
    \frac{1}{\sqrt{c \, \tau+d}} \,
    \E^{D  \frac{c z^2}{c \tau+d}  \pi \I}
    \sum_{a_2=1}^{\frac{D}{2}}
    \left[
      \rho(\gamma)
    \right]_{\frac{D}{2},a_2} \,
    {B}_{D,a_2}(z;\tau) ,
    &
    \text{for $a_1=\frac{D}{2}$.}
  \end{cases}
\end{multline}

\subsection{Harmonic Maass Forms and Elliptic Genus}
By choosing a set of coordinates on the torus,
$w_b \in \mathbb{C}/(\mathbb{Z}\oplus \tau \,
\mathbb{Z})$ for
$b=1, \cdots, \frac{D}{2}$,
we define functions $J_D(z; w_1,\dots, w_{\frac{D}{2}};\tau)$ as
\begin{equation}
  \label{J_and_C_and_H}
  \begin{aligned}[b]
    J_D(z;w_1, \dots, w_{\frac{D}{2}};\tau)
    & =
    \widehat{{C}}_{D}(z;\tau)
    -
    \sum_{a=1}^{\frac{D}{2}}
    \widehat{H}_{D,a}(w_1,\dots, w_{\frac{D}{2}} ; \tau) \,
    {B}_{D,a}(z;\tau)
    \\
    & =
    {{C}}_{D}(z;\tau)
    -
    \sum_{a=1}^{\frac{D}{2}}
    {H}_{D,a}(w_1,\dots, w_{\frac{D}{2}} ; \tau) \,
    {B}_{D,a}(z;\tau) .
  \end{aligned}
\end{equation}
Here
\begin{gather}
  \widehat{H}_{D,a}(w_1,\dots,w_{\frac{D}{2}};\tau)
  =
  \sum_{b=1}^{\frac{D}{2}}
  \left[
    \mathbf{{B}}_D(\boldsymbol{w};\tau)^{-1}
  \right]_{ab} \,
  \widehat{C}_D(w_b;\tau) ,
  \\[2mm]
  H_{D,a}(w_1,\dots,w_{\frac{D}{2}};\tau)
  =
  \sum_{b=1}^{\frac{D}{2}}
  \left[
    \mathbf{{B}}_D(\boldsymbol{w};\tau)^{-1}
  \right]_{ab} \,
  {C}_D(w_b;\tau) ,
\end{gather}
where the matrix $\mathbf{B}_D$ is defined by 
\begin{equation*}
  \left[ \mathbf{{B}}_D(\boldsymbol{w};\tau)
  \right]_{ab}
  =
  {B}_{D,b}(w_a; \tau) .
\end{equation*}
We note that
\begin{align}
  &\widehat{H}_{D,a}(w_1,\dots,w_{\frac{D}{2}};\tau)
  - H_{D,a}(w_1,\dots,w_{\frac{D}{2}};\tau)
  \nonumber
  \\
  &=
  \sum_b
  \left[
    \mathbf{{B}}_D(\boldsymbol{w};\tau)^{-1}
  \right]_{ab} \,
  \left(
    \widehat{C}_D(w_b;\tau)- {C}_D(w_b;\tau)
  \right)
  \nonumber
  \\
  &=-{1\over 2}
  \sum_b
  \left[
    \mathbf{{B}}_D(\boldsymbol{w};\tau)^{-1}
  \right]_{ab}
  \sum_c \E^{-{c-1\over D-1}\pi \I} \, q^{-{1\over 8(D-1)}} \,
  R_{{D-1\over 2},c-1}\left({-1+\tau \over
      2(D-1)};\tau\right)B_{D,c}(w_b;\tau)
  \nonumber
  \\
  &
  =-{1\over 2} \, \E^{-{a-1\over D-1}\pi \I} \,
  q^{-{1\over 8(D-1)}} \,
  R_{{D-1\over 2},a-1}\left({-1+\tau  \over 2(D-1)};\tau\right).
\end{align}
Thus $\widehat{H}_{D,a}(w_1,\dots,w_{\frac{D}{2}} ;\tau)$
is a completion of 
$H_{D,a}(w_1,\dots,w_{\frac{D}{2}};\tau)$.
By use of~\eqref{differential_R} and the fact that 
$H_{D,a}(w_1,\dots,w_{\frac{D}{2}};\tau)$ is
$\overline{\tau}$-independent, we can check that  it is annihilated by
the hyperbolic Laplacian~\eqref{Delta_phi}
\begin{equation}
  \label{H_is_harmonic_Maass}
  \Delta_{\frac{1}{2}} \widehat{H}_{D,a}(
  w_1, \dots, w_{\frac{D}{2} } ; \tau
  ) = 0 .
\end{equation}
We note that 
\begin{equation}
  \begin{gathered}
    J_D(z;w_1,\dots,w_{\frac{D}{2}} ; \tau)
    =
    \E^{- \pi \I D \frac{z^2}{\tau}} \,
    J_D\left(
      \frac{z}{\tau};
      \frac{w_1}{\tau},\dots,\frac{w_{\frac{D}{2}}}{\tau} ;
      -\frac{1}{\tau}
    \right) ,
    \\[2mm]
    \begin{aligned}
      J_D(z+1;w_1,\dots,w_{\frac{D}{2}} ; \tau)
      & =
      J_D(z;w_1,\dots,w_a+1, \dots, w_{\frac{D}{2}} ; \tau)
      \\
      & = J_D(z;w_1,\dots,w_a+\tau, \dots, w_{\frac{D}{2}} ; \tau)
      \\
      & =
      J_D(z;w_1,\dots,w_{\frac{D}{2}} ; \tau+1)
      \\
      & =
      J_D(z;w_1,\dots,w_{\frac{D}{2}} ; \tau) ,
    \end{aligned}
    \\[2mm]
    J_D(z+\tau;w_1,\dots,w_{\frac{D}{2}} ; \tau)
    =
    q^{-\frac{D}{2}} \,
    \E^{-2 \pi \I D z} \,
    J_D(z;w_1,\dots,w_{\frac{D}{2}} ; \tau) .
  \end{gathered}\label{J-prop1}
\end{equation}
By construction the function
$J_D(z;w_1,\cdots,w_{{D\over 2}};\tau)$
vanishes at $z=w_a$ for $a=1,\cdots,{D\over 2}$
\begin{equation}
  J_D(z=w_a;w_1,\cdots,w_{{D\over 2}};\tau)=0 ,
  \label{J-prop2} 
\end{equation}
and we also have 
\begin{equation}
  J_D(z=0;w_1,\cdots,w_{{D\over 2}};\tau)=1.
  \label{J-prop3} 
\end{equation}
If we choose $w_1,\dots, w_{\frac{D}{2}}$ to be half-periods
$\left\{ {1\over 2}, {1+\tau\over 2}, {\tau\over 2} \right\}$
and use the notation $\boldsymbol{w}_{(k_2,k_3,k_4)}$,
\begin{multline*}
  \boldsymbol{w}_{(k_2,k_3,k_4)}
  \\
  =
  \left\{
    w_1,\dots,w_{\frac{D}{2}}
    ~ \Big| ~
    k_2 =
    \#  \left(
      w_a=\frac{1}{2}
    \right),
    k_3
    =
    \#
    \left(
      w_a=\frac{1+\tau}{2}
    \right),
    k_4
    =
    \#
    \left(
      w_a=\frac{\tau}{2}
    \right) 
  \right\} ,
\end{multline*}
it is possible
from the above conditions~\eqref{J-prop1},
\eqref{J-prop2}, \eqref{J-prop3}
to show that 
\begin{equation}
  J_D(z; \boldsymbol{w}_{(k_2,k_3,k_4)} ; \tau)
  =
  \left(
    \frac{\theta_{10}(z;\tau)}{\theta_{10}(0;\tau)}
  \right)^{2 \, k_2}
  \,
  \left(
    \frac{\theta_{00}(z;\tau)}{\theta_{00}(0;\tau)}
  \right)^{2 \, k_3}
  \,
  \left(
    \frac{\theta_{01}(z;\tau)}{\theta_{01}(0;\tau)}
  \right)^{2 \, k_4} ,
\end{equation} 
where ${D\over 2}=k_2+k_3+k_4$ .
In this manner the functions 
$J_D(z;w_1,\dots,w_{\frac{D}{2}};\tau)$ generate the basis  
vectors of the space
$\mathbb{J}_{0,\frac{D}{2}}$ and their linear combination gives   
the elliptic genera of $CY_D$.

\subsection{Character Decomposition}
We set the Euler number of $CY_D$ to be $\chi_{CY_D}$ and consider 
a function
$Z_{CY_D}(z;\tau) - \chi_{CY_D} \,
{C}_D(z;\tau)$ which 
vanishes at $z=0$. 
Completion of this function 
$Z_{CY_D}(z;\tau) - \chi_{CY_D} \,
\widehat{{C}}_D(z;\tau)$
is a 
real analytic Jacobi form with weight-$0$ and index-$\frac{D}{2}$ and 
can be expanded in terms of
${B}_{D,a}(z;\tau)$.
This is because
a function  
$
\frac{1}{\phi_{-1,\frac{1}{2}}(z;\tau)} \,
\left(
  Z_{CY_D}(z;\tau) - \chi_{CY_D} \,
  \widehat{{C}}_D(z;\tau)
\right)$,
which is a real analytic Jacobi form with weight-$1$ and
index-$\frac{D-1}{2}\in\mathbb{Z}+\frac{1}{2}$,
can be expanded in terms of 
$\widetilde{\vartheta}_{\frac{D-1}{2},a}(z; \tau)$
as discussed in~\eqref{f_decompose_Theta_b}. 
We have
\begin{equation*}
  \frac{1}{
    \phi_{-1,\frac{1}{2}}(z;\tau)
  } \,
  \left(
    Z_{CY_D}(z;\tau) - \chi_{CY_D} \,
    \widehat{{C}}_D(z;\tau)
  \right) 
  =
  \sum_{a \mod (D-1)}
  \widehat{\Sigma}_{D,a}(\tau) \,
  \widetilde{\vartheta}_{\frac{D-1}{2},a}(z;\tau) ,
\end{equation*}
which reduces to
\begin{equation}
  \label{decomposition_complete}
  Z_{CY_D}(z;\tau) - \chi_{CY_D} \,
  \widehat{{C}}_D(z;\tau)
  =
  \sum_{a=1}^{\frac{D}{2} }
  \widehat{\Sigma}_{D,a}(\tau) \,
  {B}_{D,a}(z;\tau) .
\end{equation}
By taking the holomorphic part, we obtain a character decomposition of
the elliptic genus $Z_{CY_D}(z;\tau)$ as
\begin{equation}
  Z_{CY_D}(z;\tau)= \chi_{CY_D} \,
  {{C}}_D(z;\tau)
  +
  \sum_{a=1}^{\frac{D}{2} }
  \Sigma_{D,a}(\tau) \,
  {B}_{D,a}(z;\tau) ,
\end{equation}
where the function $\Sigma_{D,a}(\tau)$ is computed by the Fourier
integral
\begin{multline}
  \label{integral_elliptic_genus}
  {\Sigma}_{D, a}(\tau)
  \\
  =
  q^{-\frac{
      \left( a  - \frac{1}{2} \right)^2}{
      2 (D-1)
    }} \,
  \int_{z_0}^{z_0+1} \,
  \frac{1}{
    \phi_{-1,\frac{1}{2}}(z;\tau)
  } \,
  \left(
    Z_{CY_D}(z;\tau) - \chi_{CY_D} \,
    {{C}}_D(z;\tau)
  \right) \,
  \E^{-2 \pi \I \left( a - \frac{1}{2} \right) z} \,
  \mathrm{d} z ,
\end{multline}
with an  arbitrary $z_0\in\mathbb{C}$.
Since $B_{D,a}(z;\tau)$ is (the even part of) massive characters,  
 the
integral Fourier
coefficients of $\Sigma_{D,a}(\tau)$ denotes
the  multiplicities of
the massive representations.
We note that
the lowest power 
$q^{- \frac{\left( a- \frac{1}{2} \right)^2}{2 (D-1)}}$
of $\Sigma_{D,a}(\tau)$ corresponds to
the unitarity bound, and
that
it decomposes into
a sum of massless characters 
as~\eqref{unitary_bound_Q}
and~\eqref{unitary_bound_zero},
\begin{multline}
  q^{- \frac{\left( a- \frac{1}{2} \right)^2}{2 (D-1)}} \,
  {B}_{D,a}(z;\tau)
  \\
  =
  \begin{cases}
    \displaystyle
    (-1)^{\frac{D}{2}+a-1} \,
    \sum_{\varepsilon=\pm 1}
    \left[
      \ch^{\widetilde{R}}_{D,h=\frac{D}{8},Q=\varepsilon a}(z;\tau)
      +
      \ch^{\widetilde{R}}_{D,h=\frac{D}{8},Q=\varepsilon (a-1)}(z;\tau)
    \right]
    ,
    & \text{for $1 \leq a < \frac{D}{2}$},
    \\[6mm]
    \displaystyle
    -
    \left[
      \ch^{\widetilde{R}}_{D,h=\frac{D}{8},Q=\frac{D}{2}}(z;\tau)
      +
      \sum_{\varepsilon = \pm 1}
      \ch^{\widetilde{R}}_{D,h=\frac{D}{8},Q=\varepsilon \left(\frac{D}{2}-1\right)}(z;\tau)
    \right],
    &
    \text{for $a=\frac{D}{2}$}.
  \end{cases}
\end{multline}
\subsection{Poincar{\'e}--Maass Series}
Both sides of~\eqref{decomposition_complete} are  real
analytic  Jacobi forms
with weight-$0$ and index-$\frac{D}{2}$.
Because of the transformation formula~\eqref{general_modular_B} of $B_{D,a}(z;\tau)$, 
transformation law of the functions $\widehat{\Sigma}_{D,a}(\tau)$ is
determined as 
\begin{equation}
  \label{modular_Sigma}
  \widehat{\Sigma}_{D,a_1}
  \left( \gamma(\tau) \right)
  =
  \sqrt{c \, \tau + d} \,
  \sum_{a_2=1}^{\frac{D}{2} }
  \left[
    \chi(\gamma)
  \right]_{a_1, a_2} \,
  \widehat{\Sigma}_{D,a_2}(\tau),
\end{equation}
where the multiplier system $\chi(\gamma)$ is given by
\begin{multline}
  \left[ \chi \left(\gamma^{-1} \right) \right]_{a_1,a_2}
   \\
   =
   \begin{cases}
     \displaystyle
     \begin{aligned}
       &
       \frac{
         \E^{\frac{
             \left(a_1-\frac{1}{2}\right)^2
           }{D-1}
           \frac{d}{c} \pi \I
         }
       }{2 \, \sqrt{\I} \, \sqrt{c \, (D-1)}} \,
       \sum_{j=0}^{2c-1} \sum_{\varepsilon=\pm1}
       (-1)^j \, \varepsilon \,
       \E^{   (D-1) 
         \left(
           j+ \varepsilon \frac{a_2-\frac{1}{2}}{D-1}
         \right)^2
         \frac{a}{c} \pi \I
         -2 \pi \I \frac{a_1-\frac{1}{2}}{c}
         \left(
           j + \varepsilon \frac{a_2-\frac{1}{2}}{D-1}
         \right)
       } ,
       \\
       &\hspace{100mm}
       \text{for $a_2 <\frac{D}{2} $,}
     \end{aligned}
     \\[6mm]
     \displaystyle
     \begin{aligned}
       &
       \frac{
         \E^{
           \frac{\left(a_1-\frac{1}{2}\right)^2}{D-1}
           \frac{d}{c}\pi  \I}
       }{2 \, \sqrt{\I} \, \sqrt{c \, (D-1)}} \,
      \sum_{j=0}^{2c-1}
      (-1)^j
      \E^{(D-1)
        \left(j+\frac{1}{2}\right)^2
        \frac{a}{c}  \pi \I 
        -2 \pi \I \frac{a_1-\frac{1}{2}}{c}
        \left(j+\frac{1}{2} \right)
      },
      \\
      & \hspace{100mm}
      \text{for $a_2 = \frac{D}{2} $.}
    \end{aligned}
  \end{cases}
\end{multline}

In view that the fact that the basis of Jacobi forms
$J_{D}(z;w_1,\cdots,w_{{D\over 2}};\tau)$ are decomposed
as~\eqref{J_and_C_and_H}
and the elliptic genus is given by the linear combination of these
Jacobi forms, we deduce that the  functions
$\widehat{\Sigma}_{D,a}(\tau)$
are also a vector-valued harmonic Maass form as in the case of
$\widehat{H}_{D,a}(w_1,\dots,w_{\frac{D}{2}}; \tau)$
\begin{equation}
  \label{Delta_Sigma}
  \Delta_{\frac{1}{2}} \widehat{\Sigma}_{D,a}(\tau) = 0.
\end{equation}
We construct a solution of above differential
equations~\eqref{Delta_Sigma}
using the Poincar{\'e}--Maass
series $P_{D,a}(\tau)$~\cite{BrinKOno06a,BrinKOno08a}.
We note that the polar part of $P_{D,a}(\tau)$ has the form of
\begin{equation}
  \label{P_polar}
  \left.
    P_{D,a}(\tau)
  \right|_{\text{polar}}
  =
  \sum_{0 \leq n < \frac{ \left(a-\frac{1}{2}\right)^2}{2 (D-1)}}
  p_{D,a}(n) \,
  q^{n - \frac{\left(a-\frac{1}{2} \right)^2}{2 (D-1)}} .
\end{equation}
The  Fourier coefficients $p_{D,a}(n)$ of polar parts
are fixed
from~\eqref{integral_elliptic_genus} once the elliptic genus
$Z_{CY_D}(z;\tau)$ is given.
For instance, 
if the elliptic genus is given by a Jacobi form
\begin{multline}
  2^{2 k_2 -1}
  \Biggl[
    \left(
      \frac{\theta_{10}(z;\tau)}{\theta_{10}(0;\tau)}
    \right)^{2 k_2}
    \,
    \left(
      \frac{\theta_{00}(z;\tau)}{\theta_{00}(0;\tau)}
    \right)^{2 k_3}
    \,
    \left(
      \frac{\theta_{01}(z;\tau)}{\theta_{01}(0;\tau)}
    \right)^{2 k_4}
    \\
    +
    \text{other $5$ terms obtained by permutations}
  \Biggr] ,
\end{multline}
where
$k_2 + k_3+k_4=\frac{D}{2}$ and
without loss of generality
$k_2 \geq k_3 \geq k_4$,
we find that
\begin{equation}
  \label{special_p_0}
  p_{D,a}(0)
  =
  -2^{k_2-1}
  \sum_{j=2,3,4}
  \sum_{m=0}^{k_j-1}
  \frac{1}{2^{2 m}} \,
  \begin{pmatrix}
    2 \, m
    \\
    m-a+1
  \end{pmatrix} \,
  \frac{2 \, a-1}{m+a} ,
\end{equation}
where 
$\left(
  \begin{smallmatrix}
    n\\
    k
  \end{smallmatrix}
\right)$
denotes the binomial coefficient.

By use of the function $\varphi_{-h,s}^\ell(\tau)$~\eqref{define_phi}
as an eigenfunction of the hyperbolic Laplacian~\eqref{Delta_phi},
a solution of
$\Delta_{\frac{1}{2}} P_{D,a}(\tau)=0$ which transforms
like~\eqref{modular_Sigma} with the boundary  
condition~\eqref{P_polar} can be constructed in the form of
the Poincar{\'e}--Maass
series as
\begin{multline}
  P_{D,a_1}(\tau)
  = \frac{1}{\sqrt{\pi}}
  \sum_{a_2 = 1}^{\frac{D}{2} }
  \sum_{
    0 \leq m < \frac{\left(a_2-\frac{1}{2}\right)^2}{2(D-1)}
  } 
  p_{D,a_2}(m)
  \\
  \times
  \sum_{
    \gamma=\left(
      \begin{smallmatrix}
        a & b
        \\
        c & d
      \end{smallmatrix}
    \right) \in
    \Gamma_\infty \backslash \Gamma(1)
  }
  \left[
    \chi \left( \gamma^{-1} \right)
  \right]_{a_1,a_2} \,
  \frac{1}{
    \sqrt{c \, \tau+ d}
  } \,
  \varphi_{
    m - \frac{\left(a_2-\frac{1}{2}\right)^2}{2(D-1)},
    \frac{3}{4}
  }^{\frac{1}{2}} \left(
    \gamma(\tau)
  \right) ,
\end{multline}
where $\Gamma_\infty$ is the stabilizer of $\infty$,
$\Gamma_\infty
=
\left\{
  \left(
    \begin{smallmatrix}
      1 & n
      \\
      0 & 1
    \end{smallmatrix}
  \right)
  ~|~
  n\in \mathbb{Z}
\right\}$.

We have
\begin{equation}
  \widehat{\Sigma}_{D,a}(\tau)
  =
  P_{D,a}(\tau)
  +
  \Theta_{D,a}(\tau)
  .
\end{equation}
Here a weight-$1/2$ theta series
$\Theta_{D,a}(\tau)$ having 
the same
transformation law as~\eqref{modular_Sigma},
and
$\Theta_{D,a}\left( 8 \,  (D-1) \,  \tau\right)$
being a modular forms of 
$\Gamma_0\left( 64 \, (D-1)^2 \right)$ may appear in the right-hand-side of the equation.
To have a non-vanishing theta series $\Theta_{D,a}(\tau)$,
$64 \, (D-1)^2$ is divisible  either by $64 \, p^2$ with odd prime
$p$,
or by $4 \, (p \, p^\prime)^2$ with distinct odd primes $p$ and
$p^\prime$,
due to the Serre--Stark theorem~\cite{SerreStark77a}.

The holomorphic part is thus given by
\begin{equation}
  \begin{aligned}[b]
    \Sigma_{D,a}(\tau) - \Theta_{D,a}(\tau)
    & =
    \left.
      P_{D,a}(\tau)
    \right|_{\text{holomorphic}}    \\
    & =
    q^{-\frac{\left(a -\frac{1}{2}\right)^2}{2 (D-1)}}
    \sum_{n=0}^\infty
    p_{D,a}(n) \, q^n .
  \end{aligned}
\end{equation}
Following the same analysis with our previous
paper~\cite{EguchiHikami09b}
(see  Ref.~\citenum{BrinKOno08a} for a general treatment),
we can compute
the coefficients $p_{D,a}(n)$ as
\begin{equation}
  \label{p_Poincare}
  p_{D,a_1}(n)
  =
  \sum_{a_2=1}^{\frac{D}{2}}
  \sum_{
    0 \leq m <
    \frac{\left( a_2-\frac{1}{2} \right)^2}{2 (D-1)}
  }
  p_{D,a_2}(m) \,
  A_D^{(a_2,m,a_1)}(n),
\end{equation}
where
\begin{multline}
  \label{formula_for_A_D}
  A_D^{(a_2,m,a_1)}(n)
  =
  \sum_{c=1}^\infty
  \sum_{\substack{
      d \mod c
      \\
      (c,d)=1
    }}
  \left[
    \chi(\gamma^{-1})
  \right]_{a_1,a_2} \,
  \frac{2 \, \pi}{\sqrt{\I}} \,
  \left(
    \frac{
      \left(a_2-\frac{1}{2}\right)^2
      -2 \,  (D-1) \,m
    }{
      2 (D-1) \, n - 
      \left(a_1-\frac{1}{2} \right)^2
    }
  \right)^{\frac{1}{4}}
  \\
  \times
  \frac{1}{c} \,
  I_{\frac{1}{2}}
  \left(
    \frac{4 \, \pi}{c} \,
    \sqrt{
      \left(
        n - \frac{\left(a_1-\frac{1}{2}\right)^2}{2 \, (D-1)}
      \right)
      \,
      \left(
        \frac{\left(a_2-\frac{1}{2}\right)^2}{2 \, (D-1)}
        -m
      \right)
    }
  \right)
  \\
  \times
  \E^{- 2 \pi \I
    \left(
      \frac{\left(a_2-\frac{1}{2}\right)^2}{2 \, (D-1)}
      -m
    \right) \frac{a}{c}
    + 2 \pi \I
    \left(
      n - \frac{\left(a_1-\frac{1}{2}\right)^2}{2 \, (D-1)}
    \right) \frac{d}{c}
  } .
\end{multline}
We mean $a=d^{-1} \mod c$ in the summand.

Since the Fourier coefficients of theta series stay constant, they are 
negligible as compared with $p_{D,a}(n)$ which increases exponentially at large $n$.
The dominant terms at large $n$
read as
\begin{multline}
  \label{p_Poincare_dominate}
  p_{D,a_1}(n)
  \approx
  \sum_{a_2=1}^{\frac{D}{2} }
  p_{D,a_2}(0) \,
  \frac{
    2\, \pi  \,
    \left(
      \delta_{a_2, \frac{D}{2} } -2
    \right)
  }{
    \sqrt{D-1}
  } \,
  \sin
  \left(
    \frac{
      (2 \, a_1-1) \, (2 \, a_2 - 1)
    }{2 \, (D-1)} \,
    \pi
  \right)
  \\
  \times
  \left(
    \frac{
      \left( a_2-\frac{1}{2} \right)^2
    }{
      2 \, (D-1) \, n - \left(
        a_1-\frac{1}{2} \right)^2
    }
  \right)^{\frac{1}{4}} \,
  I_{\frac{1}{2}}
  \left(
    4 \, \pi \,
    \sqrt{
      \frac{\left(a_2-\frac{1}{2}\right)^2}{2 \, (D-1)} \,
      \left(
        n - \frac{\left(a_1-\frac{1}{2}\right)^2}{2 \, (D-1)}
      \right)
    }
  \right) .
\end{multline}
One finds that the exponential growth of $p_{D,a}(n)$
is determined by the maximum value
$a_2$ such that $p_{D,a_2}(0)\neq 0$.
As in the case of  hyperK{\"a}hler manifolds,
$p_{D, \frac{D}{2}}(0)$ is non-zero 
when
the elliptic genus $Z_{CY_D}(z;\tau)$ includes a Jacobi form
\begin{equation*}
  \left(
    \frac{\theta_{10}(z;\tau)}{\theta_{10}(0;\tau)}
  \right)^D
  +
  \left(
    \frac{\theta_{00}(z;\tau)}{\theta_{00}(0;\tau)}
  \right)^D
  +
  \left(
    \frac{\theta_{01}(z;\tau)}{\theta_{01}(0;\tau)}
  \right)^D .
\end{equation*}
as we see from~\eqref{special_p_0}.
In this case
 we conclude from~\eqref{p_Poincare_dominate}
that
the entropy $S_{CY_D}$ from the $U(1)$ charge-$Q$ is given by 
\begin{equation}
  \begin{aligned}[b]
    S_{CY_D}& = \log 
    \left| p_{D,Q}(n) \right|
    \\
    &   \sim
    2  \, \pi \sqrt{
      \frac{D-1}{2} \, n
      -
      \left(
        \frac{Q-\frac{1}{2}}{2}
      \right)^2
    } .
  \end{aligned}
\end{equation}
\subsection{Examples}
\subsubsection{Calabi--Yau $2$-folds $CY_2$}
It is known that in this case the $U(1)$ current algebra is enhanced to a level-1
affine $SU(2)$ current algebra, and that we have $\mathcal{N}=4$
SCA~\cite{EgucTaor86a,EgucTaor88b}.
$CY_2$ is either a complex $2$-tori or $K3$ surface.
The elliptic genus for the former vanishes.
In the latter case,
we have
$Z_{K3}(z;\tau)=2 \,  \phi_{0,1}(z;\tau)$,
and its character decomposition~\eqref{decomposition_K3}
was studied in detail in
Ref.~\citenum{EguchiHikami09a}.
Our result~\eqref{p_Poincare} for $D=2$
indeed reproduces multiplicities $A_n$~\eqref{K3_An}.

\subsubsection{Calabi--Yau $4$-folds $CY_4$}
The dimension of the space of Jacobi forms $\mathbb{J}_{0,2}$ is two, and
we set the bases of the
Jacobi forms as~\cite{EguchiHikami08a,EguchiHikami09b}
\begin{align}
  Z_{X_4^{(1)}}(z;\tau)
  & =
  16 \,
  \left[
    \left(
      \frac{\theta_{10}(z;\tau)}{\theta_{10}(0;\tau)}
    \right)^4
    +
    \left(
      \frac{\theta_{00}(z;\tau)}{\theta_{00}(0;\tau)}
    \right)^4
    +
    \left(
      \frac{\theta_{01}(z;\tau)}{\theta_{01}(0;\tau)}
    \right)^4
  \right] ,
  \\[2mm]
  Z_{X_4^{(2)}}(z;\tau)
  & =
  2 \,
  \Biggl[
    \left(
      \frac{\theta_{10}(z;\tau)}{\theta_{10}(0;\tau)} \cdot
      \frac{\theta_{00}(z;\tau)}{\theta_{00}(0;\tau)}
    \right)^2
    +
    \left(
      \frac{\theta_{00}(z;\tau)}{\theta_{00}(0;\tau)} \cdot
      \frac{\theta_{01}(z;\tau)}{\theta_{01}(0;\tau)}
    \right)^2
    \nonumber
    \\
    & \hspace{60mm}
    +
    \left(
      \frac{\theta_{01}(z;\tau)}{\theta_{01}(0;\tau)} \cdot
      \frac{\theta_{10}(z;\tau)}{\theta_{10}(0;\tau)}
    \right)^2
  \Biggr] .
\end{align}
The Eichler--Zagier bases are given by
\begin{equation}
  \begin{pmatrix}
    \left[ \phi_{0,1} \right]^2
    \\[1mm]
    \left[ \phi_{-2,1} \right]^2 \, E_4
  \end{pmatrix} 
  =
  \begin{pmatrix}
    1 & 16
    \\
    1 & - 8
  \end{pmatrix} \,
  \begin{pmatrix}
    Z_{X_4^{(1)}}(z;\tau)
    \\[1mm]
    Z_{X_4^{(2)}}(z;\tau)
  \end{pmatrix} ,
\end{equation}

In order to derive the character decomposition of these Jacobi forms,
we use~\eqref{integral_elliptic_genus} and 
find explicit results (we use $\zeta=\E^{2 \pi \I z}$ for
brevity).
\begin{multline*}
  \frac{1}{\phi_{-1,\frac{1}{2}}(z;\tau)} \,
  \left[
    Z_{X_4^{(1)}}(z;\tau) - 48 \, {C}_4(z;\tau)
  \right]
  =
  \left(
    - \zeta^{\frac{3}{2}} - 5 \, \zeta^{\frac{1}{2}}
    + 5 \, \zeta^{-\frac{1}{2}}    + \zeta^{-\frac{3}{2}} 
  \right)
  \\
  +
  \left(
    -5 \, \zeta^{\frac{5}{2}}
    - 207 \, \zeta^{\frac{3}{2}} + 790 \, \zeta^{\frac{1}{2}}
    - 790 \, \zeta^{-\frac{1}{2}}    + 207 \, \zeta^{-\frac{3}{2}} 
    + 5 \, \zeta^{-\frac{5}{2}} 
  \right) \,q
  \\
  +
  \left(
    5 \, \zeta^{\frac{7}{2}}
    +790 \, \zeta^{\frac{5}{2}}
    - 5724 \, \zeta^{\frac{3}{2}} + 13955 \, \zeta^{\frac{1}{2}}
    - 13955 \, \zeta^{-\frac{1}{2}}    + 5724 \, \zeta^{-\frac{3}{2}} 
    - 790 \, \zeta^{-\frac{5}{2}}  -
    5 \, \zeta^{- \frac{7}{2}}
  \right) \,q^2
  \\
  +
  \left(
    \zeta^{\frac{9}{2}}
    - 790 \, \zeta^{\frac{7}{2}}
    + 13955 \, \zeta^{\frac{5}{2}}
    - 65385 \, \zeta^{\frac{3}{2}}
    + 132909 \, \zeta^{\frac{1}{2}}
    \right.
    \\
    \left.
    - 132909 \, \zeta^{-\frac{1}{2}}
    + 65385 \, \zeta^{-\frac{3}{2}} 
    - 13955 \, \zeta^{-\frac{5}{2}}
    + 790 \, \zeta^{- \frac{7}{2}}
    - \zeta^{-\frac{9}{2}}
  \right) \,q^3
  +
  \cdots,
\end{multline*}
\begin{multline*}
  \frac{1}{
    \phi_{-1,\frac{1}{2}}(z;\tau)
  } \,
  \left[
    Z_{X_4^{(2)}}(z;\tau) - 6 \, {C}_4(z;\tau)
  \right]
  \\
  =
  \left( - \zeta^{\frac{1}{2}} + \zeta^{-\frac{1}{2}} \right)
  +
  \left(
    - \zeta^{\frac{5}{2}}
    +6 \, \zeta^{\frac{3}{2}} + 5 \, \zeta^{\frac{1}{2}}
    - 5 \, \zeta^{-\frac{1}{2}}    -6 \, \zeta^{-\frac{3}{2}} 
    +  \zeta^{-\frac{5}{2}} 
  \right) \, q
  \\
  +
  \left(
    \zeta^{\frac{7}{2}}
    + 5 \, \zeta^{\frac{5}{2}}
    +18  \, \zeta^{\frac{3}{2}} + 10 \, \zeta^{\frac{1}{2}}
    - 10 \, \zeta^{-\frac{1}{2}}    -18 \, \zeta^{-\frac{3}{2}} 
    -5  \, \zeta^{-\frac{5}{2}}  -
    \zeta^{- \frac{7}{2}}
  \right) \,q^2
  \\
  +
  \left(
    -5 \, \zeta^{\frac{7}{2}}
    + 10 \, \zeta^{\frac{5}{2}}
    +30  \, \zeta^{\frac{3}{2}} + 21 \, \zeta^{\frac{1}{2}}
    - 21 \, \zeta^{-\frac{1}{2}}    -30 \, \zeta^{-\frac{3}{2}} 
    -10  \, \zeta^{-\frac{5}{2}}  +
    5 \, \zeta^{- \frac{7}{2}}
  \right) \,q^3
  +\cdots .
\end{multline*}
Fourier coefficients in the above expressions
give the multiplicities of massive
representations, and
we obtain the character expansion
\begin{multline}
  \label{CY4_case_1}
  Z_{X_4^{(1)}}(z;\tau) 
  = 48 \, {C}_4(z;\tau)
  \\
  +
  q^{-\frac{1}{24}} \,  \left(
    -5 + 790 \, q + 13955 \, q^2 + 132909 \, q^3
    + 915248 \, q^4 + 5070103 \, q^5 + \cdots
  \right) \, {B}_{4,1}(z;\tau)
  \\
  +
  q^{-\frac{3}{8}} \,  \left(
    -1 -207 \, q - 5724 \, q^2 - 65385 \, q^3
    - 494145 \, q^4 - 2922021 \, q^5 - \cdots
  \right) \, {B}_{4,2}(z;\tau) ,
\end{multline}
\begin{multline}
  \label{CY4_case_2}
  Z_{X_4^{(2)}}(z;\tau) = 6 \, {C}_4(z;\tau)
  \\
  +
  q^{-\frac{1}{24}} \,  \left(
    -1 + 5 \, q + 10 \, q^2 + 21 \, q^3
    + 31 \, q^4 + 59 \, q^5 + \cdots
  \right) \, {B}_{4,1}(z;\tau)
  \\
  +
  q^{-\frac{3}{8}} \,  \left(
    6 \, q + 18 \, q^2 + 30 \, q^3
    + 60 \, q^4 + 90 \, q^5 + \cdots
  \right) \, {B}_{4,2}(z;\tau) 
  .
\end{multline}

\begin{figure}
  \centering
  \includegraphics[scale=1.0]{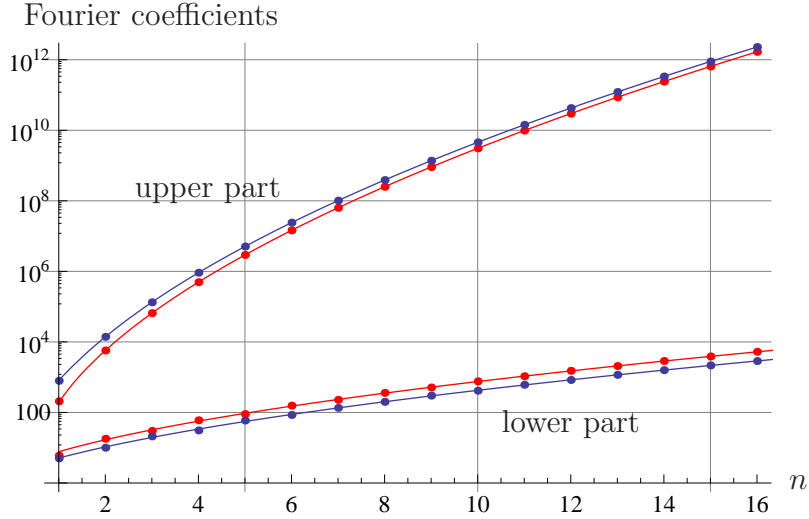}
  \caption{Shown are the 
    absolute values of the Fourier coefficients in
    the character decomposition
    of the elliptic genera for the Calabi--Yau $4$-folds.
    Blue (resp. red) dots show the exact multiplicities
    of massive representations
    associated with
    ${B}_{4,1}(z;\tau)$
    (resp.     ${B}_{4,2}(z;\tau)$)
    in~\eqref{CY4_case_1} (upper part) and~\eqref{CY4_case_2} 
    (lower part).
   Blue and red curves show the approximation by the
    dominant term~\eqref{p_Poincare_dominate}
    of the Poincar{\'e}--Maass
    series with the coefficients of the polar part  
    $(p_{4,1}(0), p_{4,2}(0))=
    (-5,-1)$ (upper part) and 
   $(-1,0)$ (lower part).
  }
  \label{fig:logCY4}
\end{figure}

In Tables~\ref{tab:CY4_case1} and~\ref{tab:CY4_case2},
we confirm numerically that the Poincar{\'e}--Maass
series~\eqref{p_Poincare} reproduce
the  above Fourier coefficients in~\eqref{CY4_case_1}
and~\eqref{CY4_case_2} accurately,
when we set
$(p_{4,1}(0), p_{4,2}(0))$ to be
$(-5,-1)$ and
$(-1,0)$, respectively.
We plot 
in Fig.~\ref{fig:logCY4} absolute values of the
Fourier coefficients,~\eqref{CY4_case_1}
and~\eqref{CY4_case_2}, together  
with the prediction of the dominant term of the Poincar{\'e}--Maass
series~\eqref{p_Poincare_dominate}.

\begin{table}
  \centering
  \begin{equation*}
    \begin{array}{rr||r|rrr}
      n & a & \text{exact} & \sum_{c=1}^1 
      & \sum_{c=1}^5 
      & \sum_{c=1}^{50}
      \\
      \hline      \hline
      1 & 1 
      & 790
      & 805.033
      & 788.286
      & 789.113
      \\
      & 2 
      & -207 
      & -187.953
      & -209.651
      & -207.055
      \\
      \hline
      4 & 1 
      & 915248
      & 914947.581
      & 915248.062
      & 915247.934
      \\
      & 2 
      & -494145
      & -494366.772
      & -494137.263
      & -494144.901
      \\
      \hline
      10 & 1 
      & 4552040952
      & 4552023847.693
      & 4552040942.289
      & 4552040950.390
      \\
      & 2 
      & -3073152762
      & -3073166628.151
      & -3073152756.944
      & -3073152762.434
      \\
      \hline
      16 & 1 
      & 2291482160295
      & 2291481818118.989
      & 2291482160303.369
      & 2291482160294.151
      \\
      & 2 
      & -1676994044877
      & -1676994336800.849
      & -1676994044862.901
      & -1676994044876.957
      \\
      \hline
    \end{array}
  \end{equation*}
  \caption{Coefficients of
    $q^{n-\frac{\left( 2 a-1 \right)^2}{24}}$ in the 
    character decomposition of $Z_{X_4^{(1)}}$~\eqref{CY4_case_1}.}
  \label{tab:CY4_case1}
\end{table}

\begin{table}
  \centering
  \begin{equation*}
    \begin{array}{rr||r|rrr}
      n & a & \text{exact} & \sum_{c=1}^1 
      & \sum_{c=1}^5 
      & \sum_{c=1}^{50}
      \\
      \hline      \hline
      1 & 1 
      & 5
      & 5.103
      & 4.797
      & 4.907
      \\
      & 2 
      & 6
      & 7.711
      & 5.815
      & 6.037
      \\
      \hline
      4 & 1 
      & 31
      & 33.770
      & 30.999
      & 31.033
      \\
      & 2 
      & 60
      & 56.661
      & 60.628
      & 60.049
      \\
      \hline
      10 & 1 
      & 414
      & 423.925
      & 413.191
      & 413.755
      \\
      & 2 
      & 762
      & 752.272
      & 762.261
      & 761.929
      \\
      \hline
      16 & 1 
      & 2865
      & 2881.769
      & 2865.012
      & 2864.833
      \\
      & 2 
      & 5256
      & 5230.583
      & 5256.358
      & 5256.004
      \\
      \hline
    \end{array}
  \end{equation*}
  \caption{Coefficients of
    $q^{n-\frac{\left( 2 a-1 \right)^2}{24}}$ in the 
    character decomposition of $Z_{X_4^{(2)}}$~\eqref{CY4_case_2}.}
  \label{tab:CY4_case2}
\end{table}

\subsubsection{Calabi--Yau $6$-folds $CY_6$}
Following the convention of 
Ref.~\citenum{EguchiHikami09b}, we set the bases of the
elliptic genera as
\begin{gather}
  Z_{X_6^{(1)}}(z;\tau)
  =
  64 \, \left[
    \left(
      \frac{\theta_{10}(z;\tau)}{\theta_{10}(0;\tau)}
    \right)^6
    +
    \left(
      \frac{\theta_{00}(z;\tau)}{\theta_{00}(0;\tau)}
    \right)^6
    +
    \left(
      \frac{\theta_{01}(z;\tau)}{\theta_{01}(0;\tau)}
    \right)^6
  \right] ,
  \\
  Z_{X_6^{(2)}}(z;\tau)
  =
  8 \, \left[
    \left(
      \frac{\theta_{10}(z;\tau)}{\theta_{10}(0;\tau)}
    \right)^4 \,
    \left(
      \frac{\theta_{00}(z;\tau)}{\theta_{00}(0;\tau)}
    \right)^2
    +
    \text{other 5 terms}
  \right] ,
  \\
  Z_{X_6^{(3)}}(z;\tau)
  =
  4 \,
  \left(
    \frac{\theta_{10}(z;\tau)}{\theta_{10}(0;\tau)}
  \right)^2 \,
  \left(
    \frac{\theta_{00}(z;\tau)}{\theta_{00}(0;\tau)}
  \right)^2 \,
  \left(
    \frac{\theta_{01}(z;\tau)}{\theta_{01}(0;\tau)}
  \right)^2 
  .
\end{gather}
We note that these bases are related to the Eichler--Zagier
bases~\cite{EichZagi85} by
\begin{equation}
  \label{Jacobi_level3}
  \begin{pmatrix}
    \left[ \phi_{0,1} \right]^3
    \\[1mm]
    \left[ \phi_{-2,1} \right]^2 \, \phi_{0,1} \, E_4
    \\[1mm]
    \left[ \phi_{-2,1} \right]^3 \, E_6
  \end{pmatrix}
  =
  \begin{pmatrix}
    1 & 24 & 96 
    \\
    1 & 0 & -48
    \\
    1 & -12 & 96
  \end{pmatrix}
  \,
  \begin{pmatrix}
    Z_{X_6^{(1)}}
    \\[1mm]
    Z_{X_6^{(2)}}
    \\[1mm]
    Z_{X_6^{(3)}}
  \end{pmatrix}
  .
\end{equation}

For these Jacobi forms, we have character decompositions as follows;
\begin{multline}
  Z_{X_6^{(1)}}(z;\tau)
  =
  192 \, {C}_6(z;\tau)
  \\
  +
  q^{-\frac{1}{40}} \,
  \left(
    - 22 + 7133 \, q + 271635 \, q^2 + 5130662 \, q^3
    + 63707417 \, q^4
    + \cdots
  \right) \, {B}_{6,1}(z;\tau)
  \\
  +
  q^{-\frac{9}{40}} \,
  \left(
    -7 - 1983 \, q -129717 \, q^2 - 2905560 \, q^3
    - 39223768 \, q^4
    - \cdots
  \right) \, {B}_{6,2}(z;\tau)
  \\
  +
  q^{-\frac{5}{8}} \,
  \left(
    -1 - 35 \, q + 26895 \, q^2 + 887110 \, q^3
    + 14389130 \, q^4
    + \cdots
  \right) \, {B}_{6,3}(z;\tau) ,
  \label{X3_case1}
\end{multline}
\begin{multline}
  Z_{X_6^{(2)}}(z;\tau)
  =
  48 \, {C}_6(z;\tau)
  \\
  +
  q^{-\frac{1}{40}} \,
  \left(
    - 9 + 197 \, q + 1599 \, q^2 + 8697 \, q^3
    + 37232 \, q^4
    + \cdots
  \right) \, {B}_{6,1}(z;\tau)
  \\
  +
  q^{-\frac{9}{40}} \,
  \left(
    -1 + 60 \, q +474 \, q^2 +2457 \, q^3
    + 10932 \, q^4
    + \cdots
  \right) \, {B}_{6,2}(z;\tau)
  \\
  +
  q^{-\frac{5}{8}} \,
  \left(
    - 26 \, q -528 \, q^2 -3954 \, q^3
    -19432 \, q^4
    - \cdots
  \right) \, {B}_{6,3}(z;\tau) ,
  \label{X3_case2}
\end{multline}
\begin{multline}
  Z_{X_6^{(3)}}(z;\tau)
  =
  4 \, {C}_6(z;\tau)
  \\
  +
  q^{-\frac{1}{40}} \,
  \left(
    - 1 +  q + 3 \, q^2 + 2 \, q^3
    + 7  \, q^4
    + \cdots
  \right) \, {B}_{6,1}(z;\tau)
  \\
  +
  q^{-\frac{9}{40}} \,
  \left(
    3 \, q + 5 \, q^2 + 9 \, q^3
    + 12 \, q^4
    + \cdots
  \right) \, {B}_{6,2}(z;\tau)
  \\
  +
  q^{-\frac{5}{8}} \,
  \left(
    2 \, q + 6 \, q^2 + 8 \, q^3
    + 14 \, q^4
    + \cdots
  \right) \, {B}_{6,3}(z;\tau) .
  \label{X3_case3}
\end{multline}

We have checked numerically that
the Poincar{\'e}--Maass series~\eqref{p_Poincare} generates the above multiplicities 
of the massive representations quite accurately,
when we set 
the polar part $(p_{6,1}(0), p_{6,2}(0), p_{6,3}(0))$ to be
$(-22,-7,-1)$,
$(-9,-1,0)$,
and
$(-1,0,0)$, respectively.
\section{Concluding Remarks}

We have computed the entropy of the Calabi--Yau $D$-folds by use of
the Poincar{\'e}--Maass series.
We have found that,
when $D$ is odd,
the entropy
coincides
with that of the hyperK{\"a}hler manifolds with
a  complex
($D-3$)-dimension.
Especially
the entropy of the Calabi--Yau 3-folds vanishes identically. 
It will be interesting to provide physical intepretation of this result.

In summary, the entropy of Calabi--Yau $D$-folds is given by 
\begin{equation}
  S_{CY_D}
  \sim
  \begin{cases}
    \displaystyle
    2 \, \pi \,
    \sqrt{
      \frac{(D-3)^2}{2 \, (D-1)} \,
      n
      -
      \left(
        \frac{D-3}{D-1} \, 
        \frac{Q-\frac{1}{2}}{2}
      \right)^2
    } ,
    &
    \text{when $D$ is odd},
    \\
    \\
    \displaystyle
    2  \, \pi \sqrt{
      \frac{D-1}{2} \, n
      -
      \left(
        \frac{Q-\frac{1}{2}}{2}
      \right)^2
    } ,
    &
    \text{when $D$ is even}.
  \end{cases}
\end{equation}

\section*{Acknowledgments}
This work is supported in part by Grant-in-Aid from the Ministry of
Education, Culture, Sports, Science and Technology of Japan.
\appendix
\section{Theta Functions}
\label{sec:Theta}
\subsection{Jacobi Theta Functions}
The Jacobi theta functions are defined by
\begin{equation}
  \begin{aligned}
    \theta_{11}(z;\tau)
    & =
    \sum_{n \in \mathbb{Z}}
    q^{\frac{1}{2} \left( n+ \frac{1}{2} \right)^2} \,
    \E^{2 \pi \I \left(n+\frac{1}{2} \right) \,
      \left( z+\frac{1}{2} \right)
    }
    ,
    \\[2mm]
    \theta_{10}(z;\tau)
    & =
    \sum_{n \in \mathbb{Z}}
    q^{\frac{1}{2} \left( n + \frac{1}{2} \right)^2} \,
    \E^{2 \pi \I \left( n+\frac{1}{2} \right) z}
    ,
    \\[2mm]
    \theta_{00} (z;\tau)
    & =
    \sum_{n \in \mathbb{Z}}
    q^{\frac{1}{2} n^2} \,
    \E^{2 \pi \I  n  z}
    ,
    \\[2mm]
    \theta_{01} (z;\tau)
    & =
    \sum_{n \in \mathbb{Z}}
    q^{\frac{1}{2} n^2} \,
    \E^{2 \pi \I n \left( z+\frac{1}{2} \right) }
    .
  \end{aligned}
\end{equation}
Throughout this paper, we set $q=\E^{2 \pi \I \tau}$ with
$\tau$ in the upper half plane,
$\tau \in \mathbb{H}$.

\subsection{Theta Function}
We define the theta function for
$D\in\mathbb{Z}$ and  $a \mod D$ by
\begin{equation}
  \vartheta_{\frac{D}{2},a}(z;\tau)
  =
  \sum_{\substack{
      n \in \mathbb{Z}
      \\
      n \equiv a \mod D
    }}
  q^{\frac{n^2}{2 D }} \, \E^{2 \pi \I n z} .
\end{equation}
We have
\begin{equation}
  \vartheta_{ \frac{D}{2} ,a}(z;\tau)
  =
  \sqrt{
    \frac{\I}{\tau}
  } \,
  \frac{1}{\sqrt{D}} \,
  \E^{-  \pi \I D \frac{z^2}{\tau}}
  \sum_{b=0}^{D-1}
  \E^{\frac{2 a b}{D} \pi \I} \,
  \vartheta_{\frac{D}{2},b}\left(
    \frac{z}{\tau} ; - \frac{1}{\tau}
  \right) .
\end{equation}
Note that
\begin{equation}
  \left(
    \vartheta_{2,1} - \vartheta_{2,-1}
  \right) (z;\tau)
  =
  - \I \, \theta_{11} (2 \, z; \tau) .
\end{equation}

For our notational convention, we introduce another set of theta series
$\widetilde{\vartheta}_{\frac{D-1}{2},a}(z;\tau)$
for $\frac{D}{2} \in \mathbb{Z}$ and
$1 \leq a \leq D-1$
\begin{equation}
  \label{define_Big_theta}
  \begin{aligned}[b]
    \widetilde{\vartheta}_{\frac{D-1}{2},a}(z;\tau)
    & =
    \E^{\frac{a-1}{D-1} \pi \I} \,
    \E^{\pi \I z} \,
    q^{\frac{1}{8(D - 1)}} \,
    \vartheta_{\frac{D-1}{2},a-1}
    \left( 
      z + \frac{-1+\tau}{2 \, (D-1)} ; \tau 
    \right)
    \\
    & =
    \sum_{n \in \mathbb{Z}}
    (-1)^n \,
    q^{\frac{1}{2 (D-1)}
      \left(
        (D-1) n + a - \frac{1}{2}
      \right)^2
    } \,
    \E^{2 \pi \I
      \left(
        (D-1) n  + a - \frac{1}{2}
      \right)
      z } .
  \end{aligned}
\end{equation}
We see that
\begin{equation}
  \widetilde{\vartheta}_{\frac{D-1}{2},a}(-z;\tau)
  =
  -
  \widetilde{\vartheta}_{\frac{D-1}{2},D-a}(z;\tau) ,
\end{equation}
and 
we have
\begin{equation}
  \begin{gathered}
    \widetilde{\vartheta}_{\frac{D-1}{2},a}(z+1;\tau)
    = - \widetilde{\vartheta}_{\frac{D-1}{2},a}(z;\tau) ,
    \\[2mm]
    \widetilde{\vartheta}_{\frac{D-1}{2},a}(z+\tau; \tau)
    =
    - q^{-\frac{D-1}{2}} \,
    \E^{-2 \pi \I (D-1) z} \,
    \widetilde{\vartheta}_{\frac{D-1}{2},a}(z;\tau) ,
    \\[2mm]
    \widetilde{\vartheta}_{\frac{D-1}{2},a}(z;\tau+1)
    =
    \begin{cases}
      \displaystyle
      \E^{\pi \I \frac{\left(a-\frac{1}{2}\right)^2}{D-1}} \,
      \widetilde{\vartheta}_{\frac{D-1}{2},a}(z;\tau),
      &
      \text{for $1\leq a \leq \frac{D}{2}$},
      \\[4mm]
      \displaystyle
      \E^{\pi \I \frac{\left(D-a - \frac{1}{2}\right)^2}{D-1}} \,
      \widetilde{\vartheta}_{\frac{D-1}{2},a}(z;\tau),
      &
      \text{for $a > \frac{D}{2}$},
    \end{cases}
    \\[2mm]
    \widetilde{\vartheta}_{\frac{D-1}{2},a_1}(z;\tau)
    =
    \sqrt{\frac{\I}{\tau}} \,
    \E^{- \pi \I(D-1) \frac{z^2}{\tau}} \,
    \frac{1}{\sqrt{D-1}} 
    \sum_{a_2=1}^{D-1}
    \E^{\frac{(2 a_1 -1) (2 a_2 -1)}{2(D-1)} \pi \I} \,
    \widetilde{\vartheta}_{\frac{D-1}{2},a_2}
    \left(
      \frac{z}{\tau} ; - \frac{1}{\tau}
    \right) .
    \label{theta-def}
  \end{gathered}
\end{equation}
Generally
for $\gamma=
\left(
  \begin{smallmatrix}
    a & b \\
    c& d
  \end{smallmatrix}
\right)
\in SL(2;\mathbb{Z})$,
we have (see, \emph{e.g.}, Refs.~\citenum{HRadema73,KOno04Book},
also Ref.~\citenum{LRozan96f} in which it is used as
surgery data in constructing the
$SO(3)$ quantum invariants of 3-manifolds)
\begin{equation}
  \label{modular_Theta}
  \widetilde{\vartheta}_{\frac{D-1}{2},a_1}\left(
    \frac{z}{c \, \tau+d} ;
    \frac{a \, \tau+b}{c \, \tau + d}
  \right)
  =
  \sqrt{c \, \tau+d}
  \,
  \E^{
    \frac{c z^2}{c \tau + d} (D-1) \pi \I}
  \sum_{a_2=1}^{D-1}
  \left[
    \rho(\gamma)
  \right]_{a_1,a_2} \,
  \widetilde{\vartheta}_{\frac{D-1}{2},a_2}(z;\tau),
\end{equation}
where $\rho(\gamma)$ is a
$(D-1)\times(D-1)$ matrix defined by
\begin{multline}
  \left[ \rho(\gamma) \right]_{a_1,a_2}
  \\
  =
  \begin{cases}
    \displaystyle
    \frac{
      \E^{\frac{d}{c} \frac{
          \left( a_2 - \frac{1}{2} \right)^2
        }{
          D-1
        } \pi \I} \,
    }{2  \, \sqrt{\I} \, \sqrt{c \, (D-1)}}
    \sum_{j=0}^{2c-1}
    (-1)^j \,
    \E^{\pi \I \frac{a}{c} (D-1)
      \left(
        j+\frac{a_1-\frac{1}{2}}{D-1}
      \right)^2
      -
      2 \pi \I \frac{a_2-\frac{1}{2}}{c} \,
      \left(
        j+\frac{a_1-\frac{1}{2}}{D-1}
      \right)
    } ,
    &
    \text{for $a_2 \leq \frac{D}{2}$},
    \\[6mm]
    \displaystyle
    -
    \frac{
      \E^{\frac{d}{c} \frac{
          \left( a_2 -D + \frac{1}{2} \right)^2
        }{
          D-1
        } \pi \I} \,
    }{2  \,\sqrt{\I} \, \sqrt{c \, (D-1)}}
    \sum_{j=0}^{2c-1}
    (-1)^j \,
    \E^{\pi \I \frac{a}{c} (D-1)
      \left(
        j+\frac{a_1-\frac{1}{2}}{D-1}
      \right)^2
      -
      2 \pi \I \frac{a_2-D+\frac{1}{2}}{c} \,
      \left(
        j+\frac{a_1-\frac{1}{2}}{D-1}
      \right)
    } ,
    &
    \text{for $a_2>\frac{D}{2}$}.
  \end{cases}
\end{multline}
(Note: the symbol $\widetilde{*}$ should not be confused with the symbol $\widehat{*}$ which signifies the completion of mock theta functions).
\section{Jacobi Forms}
\label{sec:Jacobi}
The Jacobi form with weight-$k \in \mathbb{Z}/2$ and
index-$m \in \mathbb{Z}/2$
fulfills
\begin{equation}
  \begin{gathered}
    f
    \left(
      \frac{z}{c \, \tau + d} ; \frac{a \, \tau+b}{c \, \tau+d}
    \right)
    =
    \left( c \, \tau + d \right)^k \,
    \E^{2 \pi \I m \frac{c z^2}{c \tau+d}} \,
    f(z;\tau),
    \\[2mm]
    f(z+s \, \tau + t; \tau)
    =
    (-1)^{2 m (s+t)} \,
    \E^{-2 \pi \I m (s^2 \tau + 2 s z)} \,
    f(z;\tau) ,
    \label{Jacobi-def}
  \end{gathered}
\end{equation}
where
$
\left(
  \begin{smallmatrix}
    a & b \\
    c & d
  \end{smallmatrix}
\right)
\in SL(2;\mathbb{Z})
$ and
$s,t \in \mathbb{Z}$.
The weak Jacobi form has a non-negative power of~$q$.
See Ref.~\citenum{EichZagi85} as a basic reference.

We denote $\mathbb{J}_{k,m}$ as the space of weak Jacobi forms with
weight-$k$ and index-$m$.
The space~$\mathbb{J}_{k,m}$ with even weight $k$ and integral index
$m \in \mathbb{Z}$ is generated
by~\cite{EichZagi85}
\begin{equation*}
  \left\{
    E_4(\tau) ,
    E_6(\tau) ,
    \phi_{-2,1}(z;\tau) ,
    \phi_{0,1}(z;\tau)
  \right\}  .
\end{equation*}
Namely  bases of $\mathbb{J}_{k,m}$ are
\begin{equation}
  \label{Eichler_Zagier_base}
  \left[ E_4(\tau) \right]^a \,
  \left[ E_6(\tau) \right]^b \,
  \left[ \phi_{-2,1}(z;\tau) \right]^c \,
  \left[ \phi_{0,1}(z;\tau) \right]^d,
\end{equation}
with non-negative integers
$a$, $b$, $c$, $d$ satisfying
\begin{equation*}
  \begin{aligned}
    & 4 \, a + 6 \, b - 2 \, c = k,
    &
    &c+d=m .
  \end{aligned}
\end{equation*}
Here $E_4(\tau)$ and $E_6(\tau)$ are the Eisenstein series 
defined by
\begin{equation}
  E_{2k}(\tau)
  =
  1 - \frac{4 \, k}{B_{2 k}}
  \sum_{n=1}^\infty \sigma_{2k-1}(n) \, q^n ,
\end{equation}
where $B_k$ and $\sigma_{k}(n)$ are respectively
the Bernoulli number and the
divisor function 
\begin{gather*}
  \frac{t}{\E^t -1} = \sum_{k=0}^\infty B_k \, \frac{t^k}{k!} ,
  \\[2mm]
  \sigma_k(n)
  =
  \sum_{1 \leq r | n} r^k .
\end{gather*}
The remaining two functions with index-$1$ are
\begin{align}
  \phi_{-2,1}(z;\tau)
  & =
  - \frac{
    \left[ \theta_{11}(z;\tau) \right]^2}{
    \left[ \eta(\tau) \right]^6
  } ,
  \\[2mm]
  \phi_{0,1}(z;\tau)
  & =
  4 \,
  \left[
    \left(
      \frac{\theta_{10}(z;\tau)}{\theta_{10}(0;\tau)}
    \right)^2
    +
    \left(
      \frac{\theta_{00}(z;\tau)}{\theta_{00}(0;\tau)}
    \right)^2
    +
    \left(
      \frac{\theta_{01}(z;\tau)}{\theta_{01}(0;\tau)}
    \right)^2
  \right]  .
  \label{phi01}
\end{align}

In the case 
of the index being half-odd integral,
the space $\mathbb{J}_{k,m+\frac{1}{2}}$ is isomorphic to the space
with an integral index as follows~\cite{Gritse99a,BoriLibg00a};
\begin{equation}
  \label{space_J_isomorphic}
  \begin{gathered}
    \mathbb{J}_{2k,m+\frac{1}{2}} =
    \phi_{0,\frac{3}{2}}(z;\tau) \cdot
    \mathbb{J}_{2k,m-1} ,
    \\[2mm]
    \mathbb{J}_{2k+1,m+\frac{1}{2}} =
    \phi_{-1,\frac{1}{2}}(z;\tau) \cdot
    \mathbb{J}_{2k+2,m} ,
  \end{gathered}
\end{equation}
where
$\phi_{0,\frac{3}{2}}(z;\tau)$
and
$\phi_{-1,\frac{1}{2}}(z;\tau)$
are defined by
 \begin{align}
    \phi_{0,\frac{3}{2}}(z;\tau)
    & =
    \frac{\theta_{11}(2 \, z;\tau)}{\theta_{11}(z;\tau)}
    \\
    & = 
    2 \,
    \frac{\theta_{10}(z;\tau)}{\theta_{10}(0;\tau)}
    \cdot
    \frac{\theta_{00}(z;\tau)}{\theta_{00}(0;\tau)}
    \cdot
    \frac{\theta_{01}(z;\tau)}{\theta_{01}(0;\tau)} ,
    \nonumber \\[2mm]
    \phi_{-1,\frac{1}{2}}(z;\tau)
    & =
    \frac{\I \, \theta_{11}(z;\tau)}{
      \left[ \eta(\tau) \right]^3
    } .
  \end{align}
\section{Jacobi Forms and Theta Functions}
The Jacobi form
$f(z;\tau)$ with weight-$k$ and index-$m$ can be expanded  in terms of
the theta functions
$\vartheta_{m,a}(z;\tau)$ 
(resp.
$\widetilde{\vartheta}_{m,a}(z;\tau)$)
when the index is $m \in \mathbb{Z}$
(resp. $m \in \mathbb{Z}+\frac{1}{2}$).
\begin{enumerate}
  \renewcommand{\labelenumi}{(\alph{enumi})}
\item  When $m \in \mathbb{Z}$, we have
  \begin{equation}
    \label{f_decompose_Theta_a}
    f(z;\tau)
    =
    \sum_{a \mod 2 m} \Sigma_a(\tau) \,
    \vartheta_{m,a}(z;\tau) .
  \end{equation}
  Here with an arbitrary $z_0\in \mathbb{C}$, we have
  \begin{equation}
    \label{f_Sigma_a}
    \Sigma_a(\tau)
    =
    q^{- \frac{a^2}{4 m}}
    \int_{z_0}^{z_0 +1} f(z;\tau) \,
    \E^{-2 \pi \I a z} \, \mathrm{d} z .
  \end{equation}

\item When $m \in \mathbb{Z}+\frac{1}{2}$, we have
  \begin{equation}
    \label{f_decompose_Theta_b}
    f(z;\tau)
    =
    \sum_{a \mod 2 m }
    \Sigma_a(\tau) \,
    \widetilde{\vartheta}_{ m ,a}(z;\tau) .
  \end{equation}
  Here with an arbitrary $z_0\in \mathbb{C}$, we have
  \begin{equation}
    \label{f_Sigma_b}
    \Sigma_a(\tau)
    = q^{ - \frac{1}{4 m} \left( a- \frac{1}{2} \right)^2}
    \int_{z_0}^{z_0+1}
    f(z;\tau) \,
    \E^{- 2 \pi \I  \left( a - \frac{1}{2} \right) z} \,
    \mathrm{d}z .
  \end{equation}
\end{enumerate}

The former is a standard result
(see, \emph{e.g.}, Ref.~\citenum{EichZagi85}), and the proof is as
follows.
In the case of an integral index-$m \in \mathbb{Z}$, the Jacobi form
is periodic
$f(z+1;\tau)=f(z;\tau)$, and we have the Fourier expansion
\begin{equation*}
  f(z;\tau)
  =
  \sum_{n\in \mathbb{Z}} \E^{2 \pi \I n z} \,
  q^{\frac{n^2}{4 m}} \, \Sigma_n(\tau) ,
\end{equation*}
where the prefactor is for our convention.
Then the function $\Sigma_a(\tau)$ is given by~\eqref{f_Sigma_a}.
Using $f(z+\tau;\tau) = q^{-m} \, \E^{- 4 \pi \I m z} \, f(z;\tau)$ in
the integrand, we obtain
$\Sigma_a(\tau)=\Sigma_{a+2m}(\tau)$
which  gives~\eqref{f_decompose_Theta_a}.

In the same manner,
we can prove the case (b).
When  the index $m$ is half-odd integral
$m\in \mathbb{Z}+{1\over 2}$,
the Jacobi form $f(z;\tau)$ is anti-periodic,
$f(z+1;\tau)=
-f(z;\tau)$.
Then we have the Fourier expansion
\begin{equation*}
  f(z;\tau)
  =
  \sum_{n \in \mathbb{Z}}
  \E^{2 \pi \I \left(n-\frac{1}{2} \right) z} \,
  q^{\frac{1}{4 m} \left(      n-\frac{1}{2}\right)^2} \,
  \Sigma_n(\tau) .
\end{equation*}
Here again the prefactor is chosen for our convention,
and the function $\Sigma_a(\tau)$ is defined by the Fourier
integral~\eqref{f_Sigma_b}.
Substituting
$f(z+\tau;\tau)
=
-
q^{-m} \,
\E^{-4 \pi \I m z} \, f(z;\tau)$ for~\eqref{f_Sigma_b}, we find
$\Sigma_a(\tau) = - \Sigma_{a+2m}(\tau)$
which proves~\eqref{f_decompose_Theta_b}.

\section{Hyperbolic Laplacian}
For $\tau = u  + \I \, v \in \mathbb{H}$,
we set $\Delta_\ell$ to be the hyperbolic Laplacian
\begin{equation}
  \begin{aligned}[b]
    \Delta_\ell
    & =
    - v^2 \left(
      \frac{\partial^2}{\partial u^2}
      +
      \frac{\partial^2}{\partial v^2}
    \right)
    + \I \, \ell \, v \,
    \left(
      \frac{\partial}{\partial u} + \I \,
      \frac{\partial}{\partial v}
    \right)
    \\
    & =
    -4 \,
    \left( \Im \tau \right)^{2 - \ell} 
    \frac{\partial}{\partial \tau}
    \left( \Im \tau \right)^\ell
    \frac{\partial}{\partial \overline{\tau}} .
  \end{aligned}
\end{equation}
Following Ref.~\citenum{Bruin02Book}, we set for $h>0$
\begin{equation}
  \label{define_phi}
  \varphi_{-h,s}^{\ell}(\tau)
  =
  \mathcal{M}_s^{\ell}
  \left( -4 \, \pi \, h \, \Im (\tau) \right) \,
  \E^{- 2 \,  \pi \,  \I \,  h  \, \Re(\tau)} .
\end{equation}
Here the function $\mathcal{M}_s^{\ell}(v)$ is defined by
\begin{equation*}
  \mathcal{M}_s^{\ell}(v)
  =
  \left| v \right|^{- \frac{\ell}{2}} \,
  M_{\frac{\ell}{2} \sign(v), s - \frac{1}{2}}
  \left( \left| v \right| \right) ,
\end{equation*}
where
$M_{\alpha,\beta}(z)$ is the $M$-Whittaker
function~\cite{WhittWatso27,Bruin02Book}.
The function $\varphi_{-h,s}^{\ell}(\tau)$ is an eigenfunction of
$\Delta_\ell$;
\begin{equation}
  \label{Delta_phi}
  \Delta_{\ell} \, \varphi_{-h,s}^{\ell}(\tau)
  =
  \left[
    s \, (1-s) + \frac{\ell}{2} \, \left( \frac{\ell}{2}-1 \right)
  \right] \,
  \varphi_{-h,s}^{\ell}(\tau) ,
\end{equation}
which behaves 
at $\Im \tau\to +\infty$ as
\begin{equation*}
  \varphi^{\ell}_{-h,s}(\tau)
  \sim
  \frac{\Gamma (2 \, s)}{
    \Gamma
    \left(
      \frac{\ell}{2}+s
    \right)
  } \,
  q^{-h} .
\end{equation*}

\section{Mock Theta Functions}
\label{sec:mock}

We set the Appell function~\cite{Zweg02Thesis,SemikTipunTaorm05a} for
$D \in \mathbb{Z}$,
\begin{equation}
  \label{define_f}
  f_D(u, z;\tau)
  =
  \sum_{n \in \mathbb{Z}}
  q^{\frac{D}{2} n^2} \, \frac{
    \E^{2 \pi \I D n z}}{
    1 - \E^{2 \pi \I (z - u)} \, q^n
  } .   
\end{equation}
By use of Watson's method~\cite{GWats36}, we  can check that it
transforms under the $S$-transformation as
\begin{multline}
  f_D(u, z;\tau)
  -
  \frac{
    \E^{\pi \I D \frac{u^2 - z^2}{\tau}}
  }{\tau} \,
  f_D \left(\frac{u}{\tau}, \frac{z}{\tau} ; - \frac{1}{\tau} \right)
  \\
  =
  \sum_{a=0}^{D-1}
  \vartheta_{\frac{D}{2},a}(z;\tau) \,
  \I \,
  q^{ - \frac{a^2}{2 D}} \,
  \E^{- 2 \pi \I a u} \,
  \int\limits_{\mathbb{R} - \I 0}
  \frac{
    \E^{\pi \I \tau D w^2 - 2 \pi (D u+ a \tau) w}
  }{
    1- \E^{2 \pi w}
  } \,
  \mathrm{d} w .\label{Mordell-int}
\end{multline}
Here the integration in the right hand side is the Mordell
integral~\cite{LJMorde33a}.

Following Zwegers~\cite{Zweg02Thesis}
(see  Refs.~\citenum{Zagier08a,KOno08a} for a review;
also  Refs.~\citenum{GEAndre89a,GordMcIn09a} for a classical review),
we define the completion $\widehat{f}_{D}(u,z;\tau)$ by
\begin{equation}
  \label{completion_f}
  \widehat{f}_{D}(u,z;\tau)
  =
  f_D (u,z;\tau)
  -
  \frac{1}{2}
  \sum_{a \mod D} R_{\frac{D}{2}, a} (u;\tau) \,
  \vartheta_{\frac{D}{2},a}(z;\tau) .
\end{equation}
Here the non-holomorphic partner is  defined by
\begin{equation}
  R_{\frac{D}{2},a}(u;\tau)
  =
  \sum_{\substack{
      m \in \mathbb{Z}
      \\
      m \equiv a \mod D
    }}
  \left[
    \sign \left( m+\frac{1}{2} \right)
    -
    E\left(
      \left(
        m+D \, \frac{\Im u}{\Im \tau}
      \right) \,
      \sqrt{
        \frac{2}{D} \, \Im \tau
      }
    \right)
  \right] \,
  q^{- \frac{m^2}{2 D}} \, \E^{-2 \pi \I m u} ,
  \label{R-func}
\end{equation}
where
\begin{equation*}
  E(z) = 2 \int_0^z \E^{- \pi w^2} \mathrm{d} w
  =
  1 - \erfc \left( \sqrt{\pi} \, z \right) .
\end{equation*}
Then the completion $\widehat{f}_D(u,z;\tau)$ 
has the following modular transformation
\begin{equation}
  \begin{gathered}
    \widehat{f}_D(u,z;\tau+1)
    =
    \widehat{f}_D(u,z;\tau) ,
    \\[2mm]
    \widehat{f}_D\left(
      \frac{u}{\tau} , \frac{z}{\tau} ;
      - \frac{1}{\tau}
    \right)
    =
    \tau \, 
    \E^{\pi \I D \frac{z^2-u^2}{\tau}} \,
    \widehat{f}_D(u,z;\tau) .
  \end{gathered}
\end{equation}

\section{$\mathcal{N}=4$ Superconformal Algebras}
In our previous studies on the $\mathcal{N}=4$ superconformal
algebras~\cite{EguchiHikami08a,EguchiHikami09a,EguchiHikami09b} with
central charge $c=6 \, k$,
we have used
\begin{gather}
  \label{define_B_k_a}
  B^{\mathcal{N}=4}_{k,a}(z;\tau)
  =
  \frac{\left[ \theta_{11}(z;\tau) \right]^2}{
    \left[ \eta(\tau) \right]^3
  } \,
  \frac{
    \vartheta_{k+1,a} - \vartheta_{k+1, -a}
  }{
    \vartheta_{2,1} - \vartheta_{2,-1}
  }(z;\tau) ,
  \\[2mm]
  C^{\mathcal{N}=4}_k(z;\tau)
  =
  \frac{
    \left[ \theta_{11}(z  ; \tau) \right]^2}{
    \left[ \eta(\tau) \right]^3
  } \,
  \frac{\I}{\theta_{11}(2 \, z;\tau)} \,
  \sum_{n \in \mathbb{Z}}
  q^{(k+1) n^2} \,
  \E^{4 \pi \I (k+1) n z} \,
  \frac{
    1+\E^{2 \pi \I z} \, q^n
  }{
    1 - \E^{2 \pi \I z} \, q^n
  }. 
  \label{define_C_k}
\end{gather}
Here $B^{\mathcal{N}=4}_{k,a}(z;\tau)$
with $1 \leq a \leq k$ is the basis function for the massive
characters, and $C^{\mathcal{N}=4}_k(z;\tau)$ is the massless character with
isospin-$0$.
Base functions $B^{\mathcal{N}=4}_{k,a}(z;\tau)$ are vector-valued Jacobi forms 
satisfying
\begin{equation}
  \begin{gathered}
    B^{\mathcal{N}=4}_{k,a}(z;\tau)
    = - \sqrt{\frac{\tau}{\I}} \,
    \E^{- 2  \pi  \I  k  \frac{z^2}{\tau}} \,
    \sum_{b=1}^{k}
    \sqrt{\frac{2}{k+1}} \,
    \sin \left( \frac{a \, b }{k+1} \, \pi \right) \,
    B^{\mathcal{N}=4}_{k,b} \left( \frac{z}{\tau} ; - \frac{1}{\tau} \right),
    \\[2mm]
    B^{\mathcal{N}=4}_{k,a}(z;\tau+1)
    =
    \E^{ \frac{a^2}{2(k+1)} \pi \I} \,
    B^{\mathcal{N}=4}_{k,a}(z; \tau),
    \\[2mm]
    B^{\mathcal{N}=4}_{k,a}(z+1; \tau)
    = B^{\mathcal{N}=4}_{k,a}(z; \tau) ,
    \\[2mm]
    B^{\mathcal{N}=4}_{k,a}(z+\tau ; \tau)
    =
    q^{-k} \, \E^{-4  \pi  \I  k  z} \,
    B^{\mathcal{N}=4}_{k,a}(z;\tau) ,
  \end{gathered}
\end{equation}
while the massless character  is a mock theta function satisfying
\begin{multline}
  \label{S_transform_C_k}
  C^{\mathcal{N}=4}_k(z;\tau)
  + \E^{-2 \pi \I k \frac{z^2}{\tau}} \,
  C^{\mathcal{N}=4}_k \left( \frac{z}{\tau} ; - \frac{1}{\tau} \right)
  \\
  =
  \sum_{a=0}^k
  B^{\mathcal{N}=4}_{k,a}(z;\tau) \,
  \frac{1}{2 \, (k+1)} \,
  \int\limits_{\mathbb{R}}
  \E^{\pi \I \tau \frac{w^2}{2 (k+1)}} \,
  \frac{
    \sin \left( \frac{k+1-a}{k+1} \, \pi  \right)
  }{
    \cosh \left( \frac{w}{k+1} \, \pi \right)
    + \cos \left( \frac{k+1-a}{k+1} \, \pi \right)
  } \,
  \mathrm{d} w .
\end{multline}
Completion of $C^{\mathcal{N}=4}_k(z;\tau)$ is given by
\begin{equation}
  \widehat{C}^{\mathcal{N}=4}_k(z;\tau)
  =
  C^{\mathcal{N}=4}_k(z;\tau)
  - 
  \frac{1}{
    \I \, \sqrt{2 \, (k+1)}} \,
  \sum_{a=1}^k
  R_{k,a}(0; \tau) \,
  B^{\mathcal{N}=4}_{k,a}(z;\tau) ,
\end{equation}
which is a real analytic Jacobi form satisfying
\begin{equation}
  \begin{gathered}
    \widehat{C}^{\mathcal{N}=4}_k(z;\tau)
    =
    \E^{-2  \pi   \I k  \frac{z^2}{\tau}} \,
    \widehat{C}^{\mathcal{N}=4}_k\left(\frac{z}{\tau};-\frac{1}{\tau}\right) ,
    \\[2mm]
    \widehat{C}^{\mathcal{N}=4}_k(z;\tau+1)
    =
    \widehat{C}^{\mathcal{N}=4}_k(z+1 ;\tau)
    =
    \widehat{C}^{\mathcal{N}=4}_k(z ;\tau) ,
    \\[2mm]
    \widehat{C}^{\mathcal{N}=4}_k(z+\tau ;\tau)
    =
    q^{-k} \, \E^{-4  \pi  \I  k  z} \,
    \widehat{C}^{\mathcal{N}=4}_k(z ;\tau) .
  \end{gathered}
\end{equation}

\begin{thebibliography}{10}
\providecommand{\url}[1]{\texttt{#1}}
\providecommand{\urlprefix}{URL }
\providecommand{\eprint}[2][]{\url{#2}}

\bibitem{GEAndre89a}
G.~E. Andrews, \emph{Mock theta functions}, in L.~Ehrenpreis and R.~C. Gunning,
  eds., \emph{Theta Functions --- {Bowdoin} 1987}, vol. 49 (part 2) of
  \emph{Proc. Symp. Pure Math.}, pp. 283--298, Amer. Math. Soc., Providence,
  1989.

\bibitem{BoriLibg00a}
  L.~A. Borisov and A.~Libgober,
  \emph{Elliptic genera of toric varieties and
    applications to mirror symmetry},
  \href{http://dx.doi.org/10.1007/s002220000058}{Invent. math.}
  \textbf{140},     453--485 (2000),
  \href{http://jp.arxiv.org/abs/math/9904126}{{\tt
  math/9904126}}.

\bibitem{BrinKOno06a}
  K.~Bringmann and K.~Ono,
  \emph{The $f(q)$ mock theta function conjecture and
    partition ranks},
  \href{http://dx.doi.org/10.1007/s00222-005-0493-5}{Invent.
    math.}
  \textbf{165}, 243--266 (2006).

\bibitem{BrinKOno08a}
  ---{}---{}---, \emph{Coefficients of harmonic {Maass} forms}, preprint  (2008).

\bibitem{Bruin02Book}
  J.~H. Bruinier, \emph{Borcherds Products on {$O(2,\ell)$} and {Chern} Classes
    of {Heegner} Divisors}, vol. 1780 of \emph{Lecture Notes in Mathematics},
  Springer, Berlin, 2002.

\bibitem{CvetLars98a}
  M.~Cveti\v{c} and F.~Larsen, \emph{Near horizon geometry of rotating black
    holes in five dimensions},
  \href{http://dx.doi.org/10.1016/S0550-3213(98)00604-X}{Nucl. Phys. B}
  \textbf{531}, 239--255 (1998),
  \href{http://jp.arxiv.org/abs/hep-th/9805097}{{\tt hep-th/9805097}}.

\bibitem{DijMalMooVer00a}
  R.~Dijkgraaf, J.~Maldacena, G.~Moore, and E.~Verlinde, \emph{A black hole
    {Farey} tail}, preprint  (2000),
  \href{http://jp.arxiv.org/abs/hep-th/0005003}{{\tt hep-th/0005003}}.

\bibitem{Dobre87a}
  V.~K. Dobrev, \emph{Characters of the unitarizable highest weight modules over
    the {$N=2$} superconformal algebras},
  \href{http://dx.doi.org/10.1016/0370-2693(87)90510-7}{Phys. Lett. B}
  \textbf{186}, 43--51 (1987).

\bibitem{Dyson88walk}
  F.~J. Dyson, \emph{A walk through {Ramanujan's} garden}, in \emph{Ramanujan
    Revisited}, pp. 7--28, Academic Press, Boston, 1988.

\bibitem{EguchiHikami08a}
  T.~Eguchi and K.~Hikami,
  \emph{Superconformal algebras and mock theta
    functions},
  \href{http://dx.doi.org/10.1088/1751-8113/42/30/304010}{J. Phys.
    A: Math. Theor.}
  \textbf{42}, 304010 (2009), 23 pages,
  \href{http://jp.arxiv.org/abs/0812.1151}{{\tt arXiv:0812.1151 [math-ph]}}.

\bibitem{EguchiHikami09a}
  ---{}---{}---, \emph{Superconformal algebras and mock theta functions 2.
    {Rademacher} expansion for {K3} surface}, Commun. Number Theory Phys.
  \textbf{3}, 531--554 (2009), \href{http://jp.arxiv.org/abs/0904.0911}{{\tt
      arXiv:0904.0911 [math-ph]}}.

\bibitem{EguchiHikami09b}
  ---{}---{}---, \emph{{$\mathcal{N}=4$} superconformal algebra and the entropy
    of hyper{K\"ahler} manifolds},
  \href{http://dx.doi.org/10.1007/JHEP02(2010)019}{J. High Energy Phys.}
  \textbf{2010:02}, 019 (2010), 28 pages,
  \href{http://jp.arxiv.org/abs/0909.0410}{{\tt arXiv:0909.0410 [hep-th]}}.

\bibitem{EguOogTaoYan89a}
  T.~Eguchi, H.~Ooguri, A.~Taormina, and S.-K. Yang, \emph{Superconformal
    algebras and string compactification on manifolds with {SU($n$)} holonomy},
  \href{http://dx.doi.org/10.1016/0550-3213(89)90454-9}{Nucl. Phys. B}
  \textbf{315}, 193--221 (1989).

\bibitem{EgucTaor86a}
  T.~Eguchi and A.~Taormina, \emph{Unitary representations of the {$N=4$}
    superconformal algebra},
  \href{http://dx.doi.org/10.1016/0370-2693(87)91679-0}{Phys. Lett. B}
  \textbf{196}, 75--81 (1986).

\bibitem{EgucTaor88a}
  ---{}---{}---, \emph{Character formulas for the {$N=4$} superconformal
    algebra},
  \href{http://dx.doi.org/10.1016/0370-2693(88)90778-2}{Phys. Lett.
    B}
  \textbf{200}, 315--322 (1988).

\bibitem{EgucTaor88b}
  ---{}---{}---, \emph{On the unitary representations of {$N=2$} and {$N=4$}
    superconformal algebras},
  \href{http://dx.doi.org/10.1016/0370-2693(88)90360-7}{Phys. Lett. B}
  \textbf{210}, 125--132 (1988).

\bibitem{EichZagi85}
  M.~Eichler and D.~Zagier, \emph{The Theory of {Jacobi} Forms}, vol.~55 of
  \emph{Progress in Mathematics}, Birkh{\"a}user, Boston, 1985.

\bibitem{GordMcIn09a}
  B.~Gordon and R.~J. McIntosh, \emph{A survey of classical mock theta
    functions}, preprint  (2009).

\bibitem{Gritse99a}
  V.~Gritsenko, \emph{Elliptic genus of {Calabi--Yau} manifolds and {Jacobi} and
    {Siegel} modular forms}, Algebra i Analiz \textbf{11}, 100--125 (1999),
  \href{http://jp.arxiv.org/abs/math/9906190}{{\tt math/9906190}}.

\bibitem{KawaYamaYang94a}
  T.~Kawai, Y.~Yamada, and S.-K. Yang, \emph{Elliptic genera and {$N=2$}
    superconformal field theory},
  \href{http://dx.doi.org/10.1016/0550-3213(94)90428-6}{Nucl. Phys. B}
  \textbf{414}, 191--212 (1994),
  \href{http://jp.arxiv.org/abs/hep-th/9306096}{{\tt hep-th/9306096}}.

\bibitem{Kirit88b}
  E.~B. Kiritsis, \emph{Character formulae and the structure of the
    representations of the {$N=1$}, {$N=2$} superconformal algebras},
  \href{http://dx.doi.org/10.1142/S0217751X88000795}{Int. J. Mod. Phys. A}
  \textbf{3}, 1871--1906 (1988).

\bibitem{LJMorde33a}
  L.~J. Mordell, \emph{The definite integral
    {$\int_{-\infty}^\infty \frac{\E^{a  x^2 + bx}}{\E^{cx }+d}
      \mathrm{d} x$}
    and the analytic theory of numbers},
  \href{http://dx.doi.org/10.1007/BF02547795}{Acta Math.}
  \textbf{61}, 323--360    (1933).

\bibitem{CNeum99a}
  C.~D.~D. Neumann, \emph{The elliptic genus of {Calabi--Yau} $3$- and $4$-folds,
    product formulae and generalized {Kac--Moody} algebras},
  \href{http://dx.doi.org/10.1016/S0393-0440(98)00015-1}{J. Geom. Phys.}
  \textbf{29}, 5--12 (1999),
  \href{http://jp.arxiv.org/abs/hep-th/9607029}{{\tt
  hep-th/9607029}}.

\bibitem{Odake89a}
  S.~Odake, \emph{Extension of {$N=2$} superconformal algebra and {Calabi--Yau}
    compactification},
  \href{http://dx.doi.org/10.1142/S021773238900068X}{Mod.
    Phys. Lett. A}
  \textbf{4}, 557--568 (1989).

\bibitem{Odake90b}
  ---{}---{}---, \emph{{$c=3d$} conformal algebra with extended supersymmetry},
  \href{http://dx.doi.org/10.1142/S0217732390000640}{Mod. Phys. Lett. A}
  \textbf{5}, 561--580 (1990).

\bibitem{Odake90a}
  ---{}---{}---, \emph{Character formulas of an extended superconformal algebra
    relevant to string compactification},
  \href{http://dx.doi.org/10.1142/S0217751X90000428}{Int. J. Mod. Phys. A}
  \textbf{5}, 897--914 (1990).

\bibitem{KOno04Book}
  K.~Ono, \emph{The Web of Modularity: Arithmetic of the Coefficients of Modular
    Forms and $q$-series}, no. 102 in CBMS Regional Conference Series in Math.,
  Amer. Math. Soc., Providence, 2004.

\bibitem{KOno08a}
  ---{}---{}---, \emph{Unearthing the visions of a master: harmonic {Maass} forms
    and number theory}, in D.~Jerison, B.~Mazur, T.~Mrowka, W.~Schmid, R.~P.
  Stanley, and S.-T. Yau, eds., \emph{Current Developments in Mathematics
    2008}, pp. 347--454, Intl. Press, Boston, 2009.

\bibitem{HRadema73}
  H.~Rademacher, \emph{Topics in Analytic Number Theory}, vol. 169 of
  \emph{Grund. Math. Wiss.}, Springer, New York, 1973.

\bibitem{LRozan96f}
  L.~Rozansky, \emph{Witten's invariants of rational homology spheres at prime
    values of {K} and trivial connection contribution},
  \href{http://dx.doi.org/10.1007/BF02099715}{Commun. Math. Phys.}
  \textbf{180}, 297--324 (1996),
  \href{http://jp.arxiv.org/abs/q-alg/9504015}{{\tt q-alg/9504015}}.

\bibitem{SchwiSeibe87a}
  A.~Schwimmer and N.~Seiberg, \emph{Comment on the {$N=2,3,4$} superconformal
    algebras in two dimensions},
  \href{http://dx.doi.org/10.1016/0370-2693(87)90566-1}{Phys. Lett. B}
    \textbf{184}, 191--196 (1987).

\bibitem{SemikTipunTaorm05a}
  A.~M. Semikhatov, I.~Y. Tipunin, and A.~Taormina, \emph{Higher-level {Appell}
    functions, modular transformations, and characters},
  \href{http://dx.doi.org/10.1007/s00220-004-1280-7}{Commun. Math. Phys.}
  \textbf{255}, 469--512 (2005),
  \href{http://jp.arxiv.org/abs/math/0311314}{{\tt math/0311314}}.

\bibitem{SerreStark77a}
  J.-P. Serre and H.~M. Stark, \emph{Modular forms of weight $1/2$}, in J.-P.
  Serre and D.~Zagier, eds., \emph{Modular Functions of One Variable VI}, vol.
  627 of \emph{Lecture Notes in Mathematics}, pp. 27--67, Springer, Berlin,
  1977.

\bibitem{StroVafa96a}
  A.~Strominger and C.~Vafa, \emph{Microscopic origin of the
    {Bekenstein--Hawking} entropy},
  \href{http://dx.doi.org/10.1016/0370-2693(96)00345-0}{Phys. Lett. B}
  \textbf{379}, 99--104 (1996),
  \href{http://jp.arxiv.org/abs/hep-th/9601029}{{\tt hep-th/9601029}}.

\bibitem{GWats36}
  G.~N. Watson, \emph{The final problem: an account of the mock theta functions},
  \href{http://dx.doi.org/10.1112/jlms/s1-11.1.55}{J. London Math. Soc.}
  \textbf{11}, 55--80 (1936).

\bibitem{WhittWatso27}
  E.~T. Whittaker and G.~N. Watson, \emph{A Course of Modern Analysis}, Cambridge
  Univ. Press, Cambridge, 1927, 4th ed.

\bibitem{EWitt87a}
  E.~Witten, \emph{Elliptic genera and quantum field theory},
  \href{http://dx.doi.org/10.1007/BF01208956}{Commun. Math. Phys.}
  \textbf{109}, 525--536 (1987).

\bibitem{YokoNaka00a}
  N.~Yokoi and T.~Nakatsu, 
  \emph{Three-dimensional extremal black holes and the
    {Maldacena} duality},
  \href{http://dx.doi.org/10.1143/PTP.104.439}{Prog.
    Theor. Phys.} \textbf{104}, 439--458 (2000),
  \href{http://jp.arxiv.org/abs/hep-th/9912096}{{\tt hep-th/9912096}}.

\bibitem{Zagier08a}
  D.~Zagier, \emph{Ramanujan's mock theta functions and their applications
    [d'apr{\`e}s {Zwegers} and {Bringmann}--{Ono}]}, S{\'e}minaire Bourbaki
  \textbf{986} (2006--2007).
  
\bibitem{Zweg02Thesis}
  S.~P. Zwegers, \emph{Mock Theta Functions}, Ph.D. thesis, Universiteit Utrecht
  (2002), \href{http://jp.arxiv.org/abs/0807.4834}{{\tt arXiv:0807.4834
      [math.NT]}}.
  
\end{thebibliography}

\end{document}